\documentclass[fleqn,usenatbib]{mnras}

\usepackage{newtxtext,newtxmath}
\usepackage{subfigure}

\usepackage[T1]{fontenc}
\usepackage{ae,aecompl}

\usepackage{graphicx}	
\usepackage{amsmath}	
\usepackage{amssymb}	

\title[Fingerprints of Massive Neutrinos on Cosmic Voids]{Massive Neutrinos Leave Fingerprints on Cosmic Voids}

\author[Kreisch et al.]{
Christina D. Kreisch$^{1}$\thanks{E-mail: ckreisch@astro.princeton.edu},
Alice Pisani $^{1,2}$\thanks{E-mail: apisani@astro.princeton.edu},
Carmelita Carbone$^{3,4}$,
Jia Liu $^{1}$,
\newauthor
Adam J. Hawken$^{2}$,
Elena Massara$^{5,6}$,
David N. Spergel$^{1,6}$ ,
Benjamin D. Wandelt $^{1,6,7,8}$
\\
$^{1}$Princeton University, Princeton, NJ 08544 USA\\
$^{2}$Aix-Marseille Universit\'e, CNRS/IN2P3, CPPM, Marseille, France\\
$^{3}$Universit\`a degli studi di Milano-Dipartimento di Fisica, via Celoria, 16, 20133 Milano, Italy\\
$^{4}$INAF-Osservatorio Astronomico di Brera, Via Brera, 28, 20121 Milano, Italy\\
$^{5}$Berkeley Center for Cosmological Physics, University of California, Berkeley, CA 94720 USA \\
$^{6}$Center for Computational Astrophysics, Flatiron Institute, 162 5th Avenue, New York, NY 10010 USA \\
$^{7}$Institut d'Astrophysique de Paris, 98bis Boulevard Arago, 75014 Paris, France \\
$^{8}$Sorbonne Universit\'es, Institut Lagrange de Paris, 98 bis Boulevard Arago, 75014 Paris, France
}

\pubyear{2018}

\newcommand{\hmone}{${h}^{-1}$}

\hypersetup{draft}
\begin{document}
\label{firstpage}
\pagerange{\pageref{firstpage}--\pageref{lastpage}}
\maketitle

\begin{abstract}
{Do void statistics contain information beyond the tracer 2-point correlation function? Yes! As we vary the sum of the neutrino masses, we find void statistics contain information absent when using just tracer 2-point statistics.} Massive neutrinos uniquely affect cosmic voids. We explore their impact on void clustering using both the \texttt{DEMNUni} and \texttt{MassiveNuS} simulations. For voids, neutrino effects depend on the observed void tracers. As the neutrino mass increases, the number of small voids traced by cold dark matter particles increases and the number of large voids decreases. Surprisingly, when massive, highly biased, halos are used as tracers, we find the opposite effect. The scale at which voids cluster, as well as the void correlation, is similarly sensitive to {the sum of neutrino masses and} the tracers. This scale dependent trend is not due to simulation volume or halo density. The interplay of these signatures in the void abundance and clustering leaves a distinct fingerprint that could be detected with observations and potentially help break degeneracies between different cosmological parameters. This paper paves the way to exploit cosmic voids in future surveys to constrain the mass of neutrinos.
\end{abstract}



\section{Introduction}

{Can the underdense regions in our universe reveal information inaccessible to the dense regions?} The cosmic web \citep{Bond1996} is a powerful tool to constrain neutrino properties. Cosmic voids are large (typically $10-100\,h^{-1}\mathrm{Mpc}$) underdense regions of the cosmic web that have undergone minimal virialization and are dominated by inward or outward bulk flows \citep{Gregory1978,Shandarin2010,Falck2015,Ramachandra2017}. In contrast to halos, which have undergone non-linear growth that can wash out primordial information, voids offer a pristine environment to study cosmology. As such, voids are a complementary probe to measurements of the cosmic microwave background and galaxy clustering and can help break existing degeneracies between cosmological parameters, thus becoming increasingly popular to study with both simulations and observations \citep[see e.g.][and references therein]{Ryden1995,Goldberg2004,Colberg2008,Viel2008,VanWeygaertErwinPlaten2009,Sheth2003,Chan2014,Hamaus2014,Sutter2014,Sutter2014a,Hamaus2015a,Szapudi2015,Qin2017,Alonso2017,Pollina2018,voidwhitepaper}.

The discovery of neutrino oscillations demonstrates that at least two neutrino families must have a nonzero mass \citep{Becker-Szendy1992,Fukuda1998,Ahmed2004}, evidence for beyond the standard model physics. Cosmological observables provide stringent upper bounds on the sum of neutrino masses, $\sum m_\nu$ \citep[see e.g.][]{PlanckCollaboration2018}, and may soon determine the last missing parameter in the standard model.

At linear order, neutrinos do not cluster on scales smaller than their free-streaming length, which is a function of the mass $m_\nu$ of the single neutrino species \citep{Lesgourgues2006a}. For example, neutrinos have free-streaming lengths of $130\,h^{-1}\mathrm{Mpc}$ and $39\,h^{-1}\mathrm{Mpc}$ for $\sum m_\nu = 0.06\,\mathrm{eV}$ and $\sum m_\nu = 0.6\,\mathrm{eV}$ (assuming 3 degenerate neutrino species), respectively. Neutrino free-streaming scales for $\sum m_\nu$ of interest thus fall within the range of typical void sizes, making voids an interesting tool for studying neutrinos.

Voids are sensitive to a number of effects, such as: redshift space distortions and the relative growth rate of cosmic structure \citep[e.g.][]{Paz2013,Hamaus2016,Achitouv2017,Hamaus2017,Hawken2017}, Alcock-Paczy{\'n}ski distortions \citep[e.g.][]{Alcock1979,Lavaux2012a,Sutter2012a,Sutter2014c,Hamaus2014b,Hamaus2016,Mao2017,Achitouv2018}, weak gravitational lensing \citep[e.g.][]{Melchior2014,Clampitt2015,Sanchez2016,Chantavat2017}, baryon acoustic oscillations \citep{Kitaura2016}, and the integrated Sachs-Wolfe effect \citep[e.g.][]{Granett2008,Ili2013,Kovacs2015,Kovacs2016,Nadathur2016,Naidoo2016,Cai2017,Kovacs2017}.

Voids offer an environment with unique sensitivity to signatures of physics beyond the standard model. They are one of the best observables to probe theories of gravity \citep{Odrzywoek2009,Li2012,Clampitt2013,Cai2014,Gibbons2014,Zivick2014,Barreira2015,Hamaus2016,Baldi2016} and dark energy \citep{Lee2009,Bos2012,Lavaux2012a,Sutter2014d,Pisani,Pollina2015}.

Since voids are under-dense in matter, they are particularly sensitive to the effects of diffuse components in the universe like radiation and dark energy. For this reason, voids offer an appealing, new  avenue to constrain neutrino properties. \citet{Villaescusa-Navarro2013a} studied how massive neutrinos affect voids at high redshifts with Ly$\alpha$ forest analyses using hydrodynamical simulations \citep[see also][]{Krolewski2017}. \citet{Massara2015} focused on how neutrinos affect void abundance, density profiles, ellipticities, the correlation function, and velocity profiles with N-body simulations that included massive neutrinos as an additional collisionless particle component. \citet{Banerjee2016} observed that neutrinos affect the scale-dependent void bias for voids traced by the CDM particle field. They use a spherical void finder and a small volume simulation (700~$h^{-1}$Mpc box length). In recent data analyses voids have been found using finders that do not assume spherical voids \citep[e.g.][]{Hamaus2017,Pollina2017}. It is interesting to analyze the effects of neutrinos on voids with non-spherical shapes, such as in \citet{Massara2015}, which have the advantage of closely following the cosmic web pattern. Work such as \citet{Hamaus2014a} analyzed void power spectra without discussion of neutrinos. Thus far, the effect of neutrinos on voids has not been considered in depth without assuming spherical voids, and their effect on voids traced by halos is especially unexplored. Previous simulations with massive neutrinos did not have the volume and resolution to explore the effect of neutrinos on voids derived from the halo distribution and Halo Occupation Distribution (HOD) mocks \citep[see e.g.][]{Massara2015}.

We use N-body simulations with densities and volumes large enough to distinguish the effects neutrinos have on voids derived from the halo distribution and on voids derived from the particle distribution. { Both the void size distribution and clustering respond to $\sum m_\nu$. We uncover and resolve the apparent paradox that voids found in the halo field respond in the opposite manner to $\sum m_\nu$ than voids found in the particle field. The impact of $\sum m_\nu$ on voids changes sign as a function of halo bias. $\sum m_\nu$'s sign dependent impact on void size and clustering does not occur for other cosmological parameters such as $\sigma_8$. The void exclusion scale shifts in response to $\sum m_\nu$, as well, a scale-dependent response unique to voids. The response of voids to $\sum m_\nu$ is thus novel-- neutrinos leave unique fingerprints on voids.}

The paper is organized as follows. In \S\ref{sec:sim+nbody} we describe the two sets of massive neutrino simulations used in this work, the Dark Energy and Massive Neutrino Universe Project (\texttt{DEMNUni}) and the Cosmological Massive Neutrino Simulations (\texttt{MassiveNuS}), as well as the void finder used to build our void catalog. We show how neutrinos impact voids in \S\ref{sec:results} and discuss these results in \S\ref{sec:discussion}. We conclude and discuss application to future surveys in \S\ref{sec:conclusions}.

\section{Simulations and void finder}
\label{sec:sim+nbody}

In this work, we use two sets of massive neutrino simulations: the Dark Energy and Massive Neutrino Universe \citep[\texttt{DEMNUni},][]{Carbone2016,Castorina2016}, and the Cosmological Massive Neutrino Simulations \citep[\texttt{MassiveNuS}\footnote{The \texttt{MassiveNuS} data products, including snapshots, halo catalogues, merger trees, and galaxy and CMB lensing convergence maps, are publicly available at \url{http://ColumbiaLensing.org}.},][]{Liu2017}. We isolate the effects of $\sum m_\nu$ by comparing the large volume DEMNUni simulations (2~$h^{-1}$Gpc box length, 2048$^3$ CDM particles plus 2048$^3$ $\nu$ particles) with the smaller but more highly resolved \texttt{MassiveNuS} simulations (512~$h^{-1}$Mpc box length, 1024$^3$ CDM particles-- i.e. eight times higher resolution than \texttt{DEMNUni} but 60 times smaller in volume). We focus our analysis on the simulation snapshots at $z=0$.

Comparing how neutrinos affect voids for different tracers is imperative when looking towards constraining the sum of neutrino masses with upcoming surveys \citep[see e.g.][for $\sum m_\nu$ constraint sources in galaxy surveys]{Boyle2018}. Surveys observe galaxies, which are biased tracers of the CDM fluctuations \citep{Villaescusa-Navarro2013,Castorina2013}, and void properties are sensitive to the tracer used to build the void catalog \citep{Pollina2015,Pollina2017}. We rely on the optimal features of both simulations to be sensitive to neutrino effects at different scales, show consistency, check that our results are physical, and robustly test the sensitivity of our results to simulation design (see Appendix \ref{sec:vol_res} for volume and resolution tests). The small volume and high resolution of \texttt{MassiveNuS} causes these simulations to be dominated by small voids, capturing the small scale impacts of $\sum m_\nu$, whereas the large volume of the \texttt{DEMNUni} simulations captures large scale effects. \texttt{MassiveNuS}'s high resolution enables the use of halos above a minimum mass $M_\mathrm{min}=3 \times 10^{11}\,h^{-1}M_\odot$ whereas \texttt{DEMNUni}'s minimum halo mass is $M_\mathrm{min}=2.5 \times 10^{12}\,h^{-1}M_\odot$, making \texttt{MassiveNuS} halos less biased than \texttt{DEMNUni}. The two simulations also use different methods to capture the effect of massive neutrinos-- \texttt{DEMNUni} neutrinos are treated as particles and \texttt{MassiveNuS} neutrinos use a fast linear response algorithm \citep{Ali-Haimoud2013}. { Due to the impact of mass resolution on the halo catalogs, we now denote the \texttt{DEMNUni} simulations as `low-res' and the \texttt{MassiveNuS} simulations as `high-res' throughout our analysis. We note, however, that both the simulation mass resolution and the simulation volume impact the size of the voids:
\begin{itemize}
    \item For a fixed simulation volume: a lower mass resolution simulation has more large voids than a higher mass resolution simulation. Conceptually, this can be thought of in terms of the simulation's minimum halo mass-- a larger minimum halo mass yields larger voids. We describe this further in \autoref{sec:tracer_bias} and \autoref{sec:discussion}.
    \item For a fixed simulation mass resolution: the size of the largest void is larger for the simulation with larger volume\footnote{{An important caveat to this is if voids have a maximum physical scale and if both simulations are large enough to capture this physical scale. In this case, the size of the largest void in each simulation (even if the simulations have different volumes) would be the same. In our work, however, our simulations only contain voids in size up to $\approx 100\,h^{-1}\mathrm{Mpc}$, and voids of this size have been observed \citep[see e.g. Figure 1 in][]{Hamaus2017}.}}. For example, in our work, the maximum void radius in the \texttt{DEMNUni} massless neutrino CDM field is $79\,h^{-1}\mathrm{Mpc}$, whereas the maximum void radius in the \texttt{MassiveNuS} massless neutrino CDM field is $37\,h^{-1}\mathrm{Mpc}$. Further, the void abundance smoothly decreases as a function of void size. Thus, there will be a greater number of the small simulation's largest voids in the larger simulation. This, then, causes the larger simulation to have better uncertainties for measurements relating to large voids since the larger simulation has more large voids than the smaller simulation.
\end{itemize}
Therefore, the \texttt{DEMNUni} simulations contain more large voids, and larger voids in general, than the \texttt{MassiveNuS} simulations due to both \texttt{DEMNUni}'s lower mass resolution and larger volume.}

\begin{figure}
\begin{center}
\includegraphics[width=0.5\textwidth]{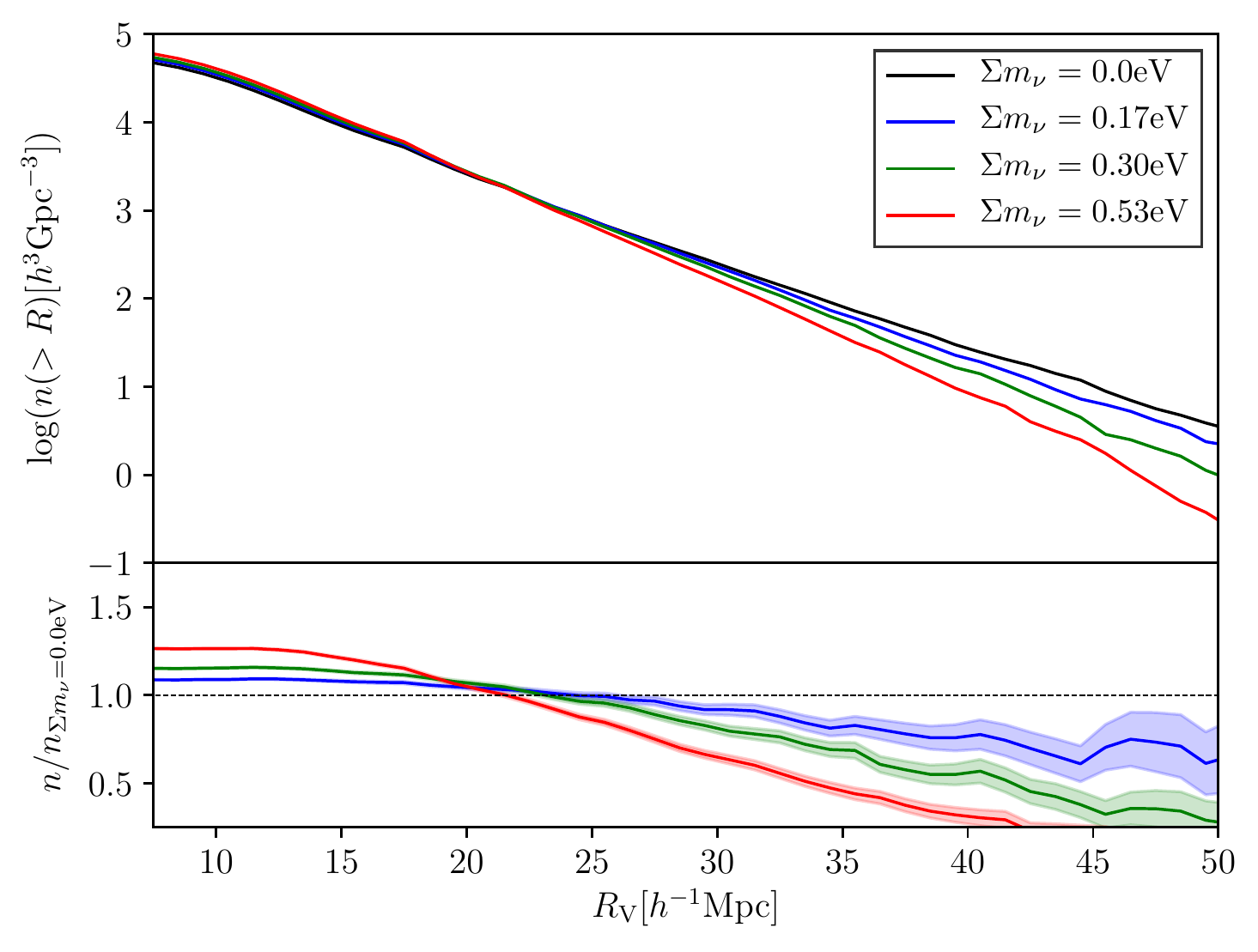}
\caption{Void abundance in the sub-sampled cold dark matter field of the \texttt{DEMNUni} simulation. Colors denote the sum of neutrino masses used in each simulation. The bottom panel shows the ratio between void number densities (with uncertainties) for different $\sum m_\nu$ values and the number density in the massless neutrino case. Increasing $\sum m_\nu$ increases the number of small voids and decreases the number of large voids derived from the particle field. All abundance plots are cut at $\sim 2$ times the mean particle separation in the simulation and where voids are so large that there are too few voids for informative uncertainties. All figures are for $z=0$.}
\label{fig:demnuni_CDM_void_abundance}
\end{center}
\end{figure}

\begin{figure}
\begin{center}
\includegraphics[width=0.5\textwidth]{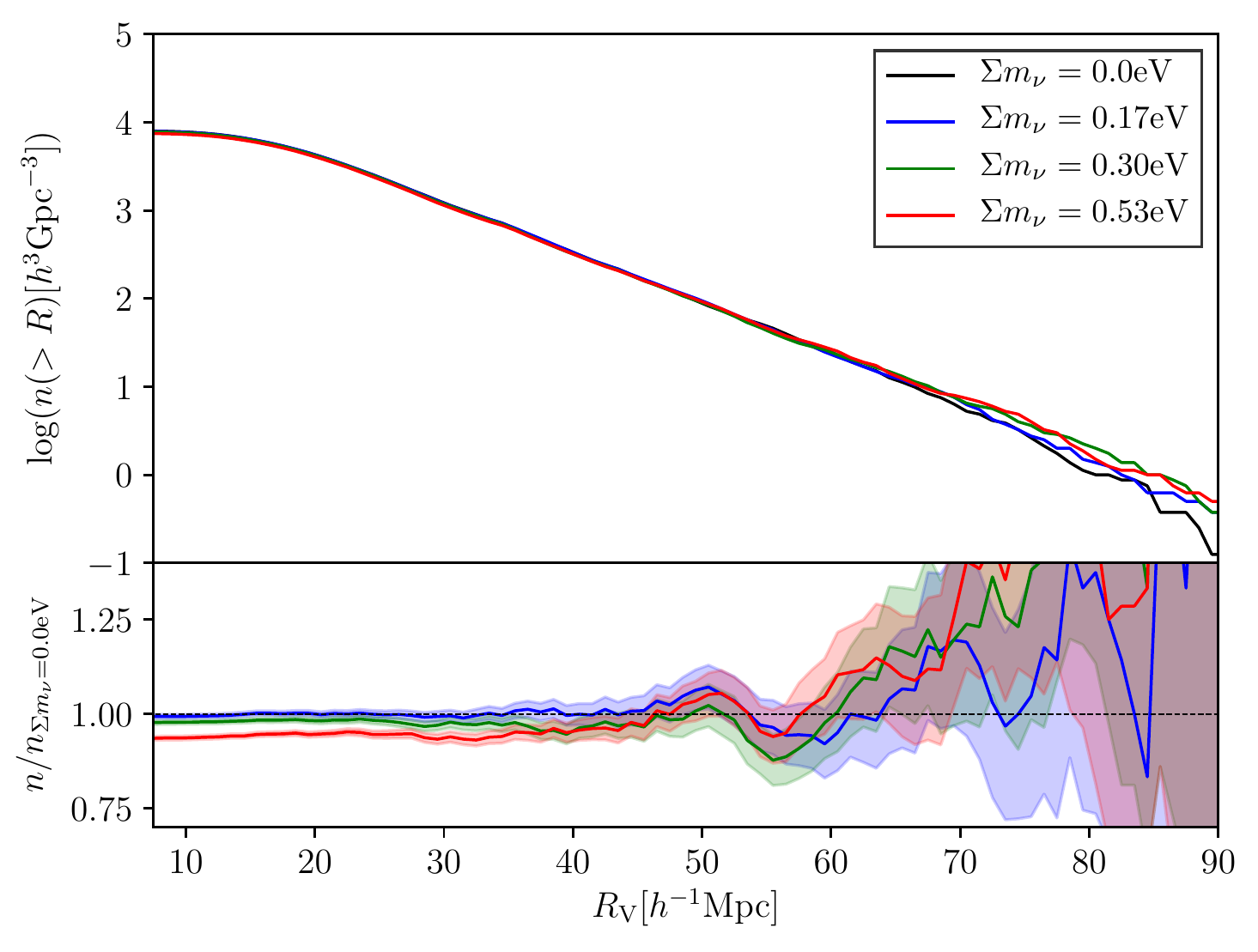}
\caption{Void abundance in the halo field of the `low-res' simulation. Colors denote the sum of neutrino masses used in each simulation. The bottom panel shows the ratio between void number density with uncertainties for the different $\sum m_\nu$ values and the number density in the massless neutrino case. Increasing $\sum m_\nu$ decreases the number of small voids and increases the number of large voids derived from the halo field.}
\label{fig:demnuni_halo_ratio}
\end{center}
\end{figure}

The sum of neutrino masses $\sum m_\nu$ is varied in each simulation suite with other cosmological parameters kept fixed. The \texttt{DEMNUni} simulations assume a baseline cosmology according to the Planck results \citep{PlanckCollaboration2014}, with $h = 0.67$, $n_s = 0.96$, $A_s = 2.1265 \times 10^{-9}$, $\Omega_m = 0.32$, and $\Omega_b = 0.05$. The relative energy densities of cold dark matter $\Omega_c$ (and neutrinos, $\Omega_\nu$) vary for each model as $\Omega_c =0.27$,  $0.2659$,  $0.2628$ and  $0.2573$, for $\sum m_\nu = 0$, $0.17$, $0.30$ and $0.53$ eV, respectively. In the considered cases, since $A_s$ is fixed while varying the neutrino mass, the simulations with massive neutrinos have a lower value of $\sigma_8$ with respect to the massless neutrino $\Lambda$CDM case. We use the three fiducial models of \texttt{MassiveNuS} in this work, where $\sum m_\nu=0$, 0.1, 0.6~eV and all other parameters are held constant at $A_s$=2.1$\times 10^{-9}$, $\Omega_m$=0.3, $h$=0.7, $n_s$=0.97, $w$=$-1$, and $\Omega_b$=0.05.

We use the public void finder \texttt{VIDE}\footnote{\url{https://bitbucket.org/cosmicvoids/vide_public}, version most recently updated on $2017-11-27$.} to locate voids in the simulations \citep{Sutter}. Because the void finder runs on a tracer distribution and uses the position of these objects, we can find voids from both the halo distribution (in this work we use the friends-of-friends (FoF) catalogs) and the CDM particle distribution. For the latter, running the void finding procedure on a large number of CDM particles (e.g. directly on the~$2048^3$~particles) is computationally expensive. We thus subsampled the CDM particle field to $1.5\%$ of the original particle number for both \texttt{DEMNUni} and \texttt{MassiveNuS}. See \citet{Sutter2014b} for a discussion on how subsampling the tracer distribution affects voids. We note that for the \texttt{DEMNUni} subsampling this corresponds roughly to $505^3$ particles, which is comparable to the CDM particle number density in the work done by \citet{Massara2015}. Throughout the paper we refer to the subsampled CDM particle field simply as ``CDM particles". We do not subsample the halo field unless specified. See Appendix \ref{sec:sim_details} for more information on the simulations and void finder.

\section{Results}
\label{sec:results}

The sum of neutrino masses affects both the number of voids and the void bias. As the sum of neutrino masses increases, there are fewer large voids and more small voids seen in the CDM field. However, if we use halos as tracers there are more large voids and fewer small voids. The total number of voids changes, as well (see Section \ref{sec:abundance}). Neutrinos affect how voids cluster and produce a strong scale dependent trend-- this is a distinctive feature (see Section \ref{sec:powerspectra}).

We note that we have also analyzed void catalogs built from the mock HOD\footnote{Contact Adam Hawken for the HOD code at adamhawken@gmail.com} galaxy catalog obtained from the DEMNUni simulations. The HOD's are built using the model described in \citet{Zheng2005}, and the luminosity dependence is described in \citet{DeLaTorre2013}. Results for the HOD catalogs are consistent with those obtained for the halo field. From now on we focus our analysis only on void catalogs extracted from the CDM and halo fields.

\subsection{Void abundance}
\label{sec:abundance}

The impact of $\sum m_\nu$ on the void abundance, i.e. the void size function, depends upon the tracer. In \autoref{fig:demnuni_CDM_void_abundance} and \autoref{fig:demnuni_halo_ratio} we show the void abundances derived from the subsampled CDM distribution and the halo distribution, respectively, for the `low-res' simulation. All abundance plots have Poisson uncertainties.

\begin{figure}
\begin{center}
\includegraphics[width=0.5\textwidth]{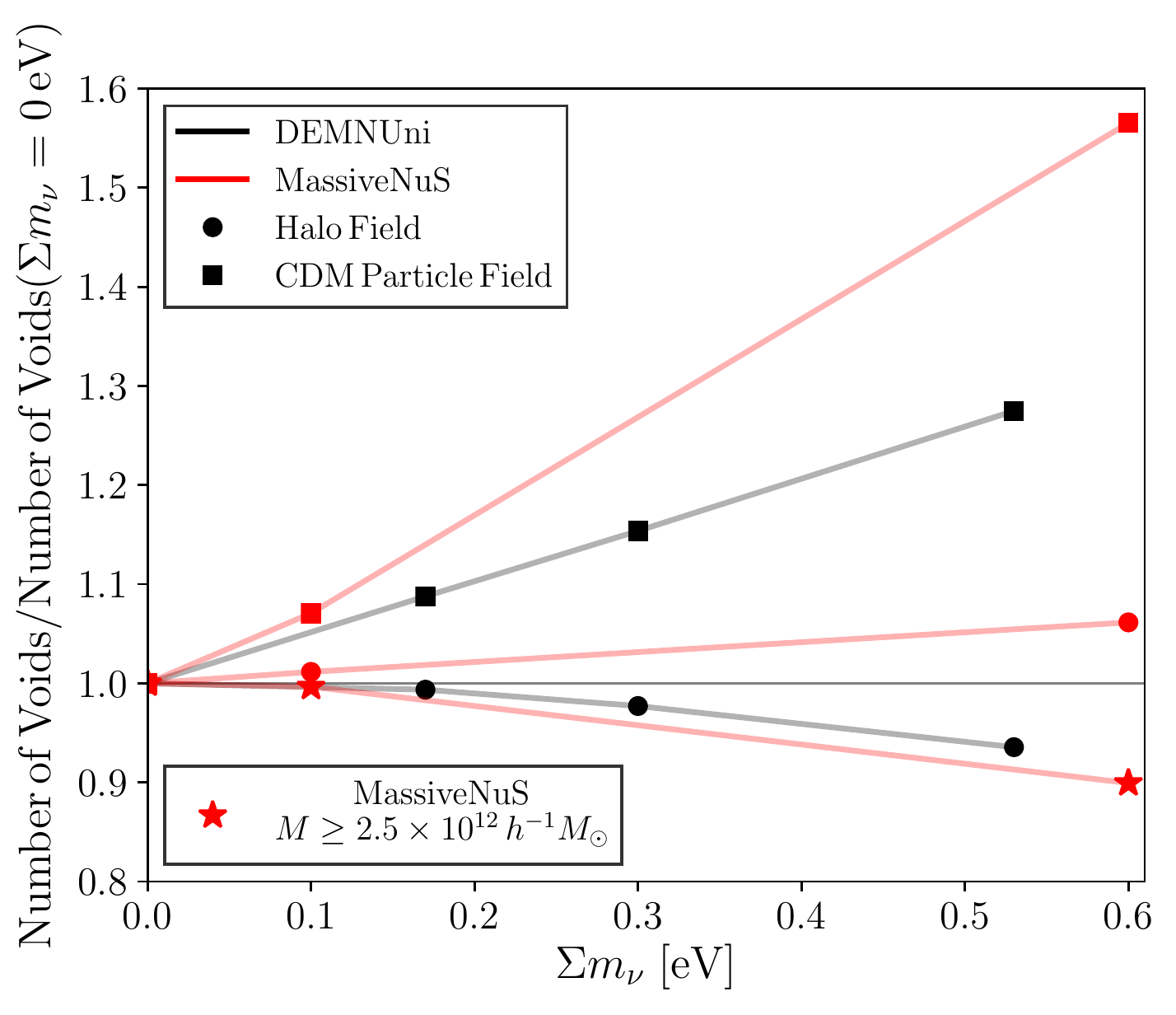}
\caption{The total number of voids for each simulation and each tracer as a function of the sum of neutrino masses. The number is normalized to the number of voids in the simulation when $\sum m_\nu=0\,\mathrm{eV}$. The normalization values are 63822, 441174, 4765, and 22337 for the `low-res' (\texttt{DEMNUni}) halos case, `low-res' (\texttt{DEMNUni}) CDM case, `high-res' (\texttt{MassiveNuS}) halos case, and `high-res' (\texttt{MassiveNuS}) CDM case, respectively. The total number of voids increases with $\sum m_\nu$ for voids traced by cold dark matter and decreases with $\sum m_\nu$ for voids traced by halos with a high mass threshold. The range of $\sum m_\nu$ spans values covered by the simulations in our analysis.}
\label{fig:void_number}
\end{center}
\end{figure}

The trend with $\sum m_\nu$ for the void abundance derived from the halo distribution is inverted relative to that derived from the CDM particle field. The void abundance derived from the CDM field shows that increasing $\sum m_\nu$ increases the number of small voids and decreases the number of large voids. Our findings are consistent with \citet{Massara2015}'s results based on a simulation with lower volume and mass resolution than our `low-res' simulation. Conversely, for the void abundance derived from the halo distribution (see \autoref{fig:demnuni_halo_ratio}) increasing $\sum m_\nu$ \textit{decreases} the number of small voids and \textit{increases} the number of large voids, although the magnitude of the effect is lower in absolute value than in the CDM case. As explained in Appendix \ref{sec:vol_res}, although the number density of the tracers changes when changing $\sum m_\nu$, the number density is not the origin of the opposite trends observed in the different void abundance plots.

Previous simulations lacked a sufficient combination of volume and mass resolution to investigate the void abundance derived from the halo field in detail and so were unable to discriminate between these two different trends in the void statistics \citep[see e.g. Section 5 of][whose simulations had $512^3$ CDM particles, $512^3$ neutrinos, and a $500\,h^{-1}\mathrm{Mpc}$ box length]{Massara2015}.

Varying $\sum m_\nu$ not only impacts the void abundance but also the total number of voids, as expected. In \autoref{fig:void_number} we show the total number of voids in the `low-res' and `high-res' simulations derived from both the halo distribution and CDM particle distribution as a function of $\sum m_\nu$. For voids derived from the CDM distribution, the total number of voids increases as $\sum m_\nu$ increases. There are more small voids and less large voids for the CDM case as $\sum m_\nu$ increases. The simulation volume is kept fixed, so overall there is a larger total number of voids that fill the volume.

For the halo case, the `low-res' and `high-res' simulations show opposite behavior for the total number of voids as a function of $\sum m_\nu$. Increasing $\sum m_\nu$ decreases the total number of `low-res' voids derived from the halo field. This occurs because increasing $\sum m_\nu$ decreases the number of small voids and increases the number of large voids in the `low-res' halo case, so there must be a lower (with respect to the massless neutrino case) total number of voids to fill the simulation volume. For the `high-res' simulations, the number of voids increases with $\sum m_\nu$ in both the halo and CDM cases. The `high-res' simulations have a smaller volume than the `low-res' simulation, yielding a smaller total number of voids and, thus, larger uncertainties in the void abundance than the `low-res' simulation. Nonetheless, the `high-res' void abundances for both the halo case and CDM case appear to be consistent with the trends seen in the `low-res' CDM case. Thus, since the `high-res' simulation has more small voids for both the halo case and CDM case for nonzero $\sum m_\nu$ relative to the massless case, the total number of voids must also increase with $\sum m_\nu$ in both cases. We include the `high-res' void abundances in Appendix \ref{sec:massivenus_abundance}. The `high-res' simulation has halos with smaller masses than the `low-res' simulation, and thresholding the halo mass in the `high-res' simulation to match that of the `low-res' simulation gives concordance between the two simulations for the total number of voids traced by halos.

\subsection{Power Spectra \& Correlation Functions}
\label{sec:powerspectra}

\begin{figure*}%
\centering
\subfigure[][]{%
\centering
\label{fig:PmmDEMNUni}%
\includegraphics[width=0.4\textwidth]{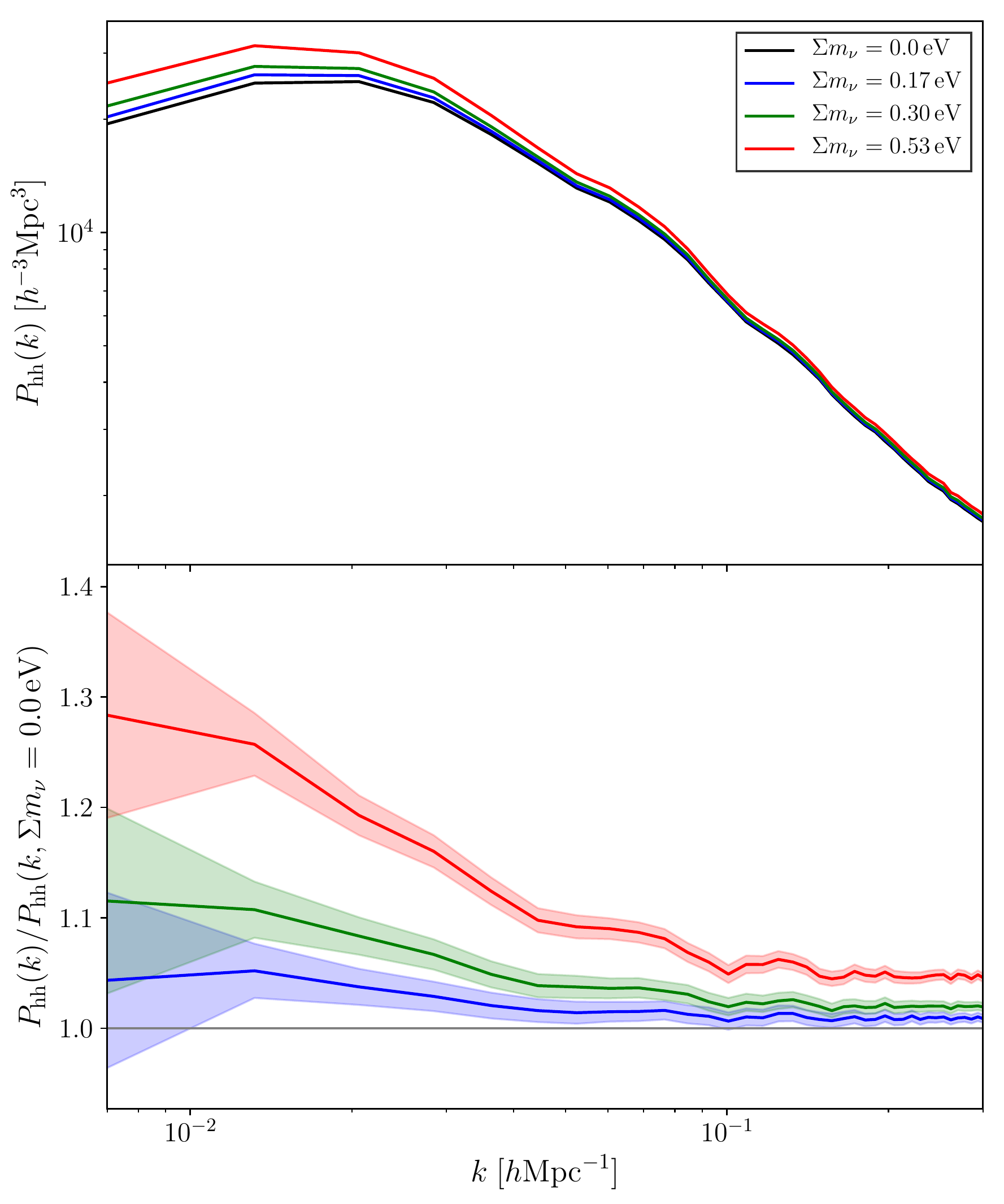}}%
\hspace{16pt}%
\subfigure[][]{%
\centering
\label{fig:PvvDEMNUni}%
\includegraphics[width=0.4\textwidth]{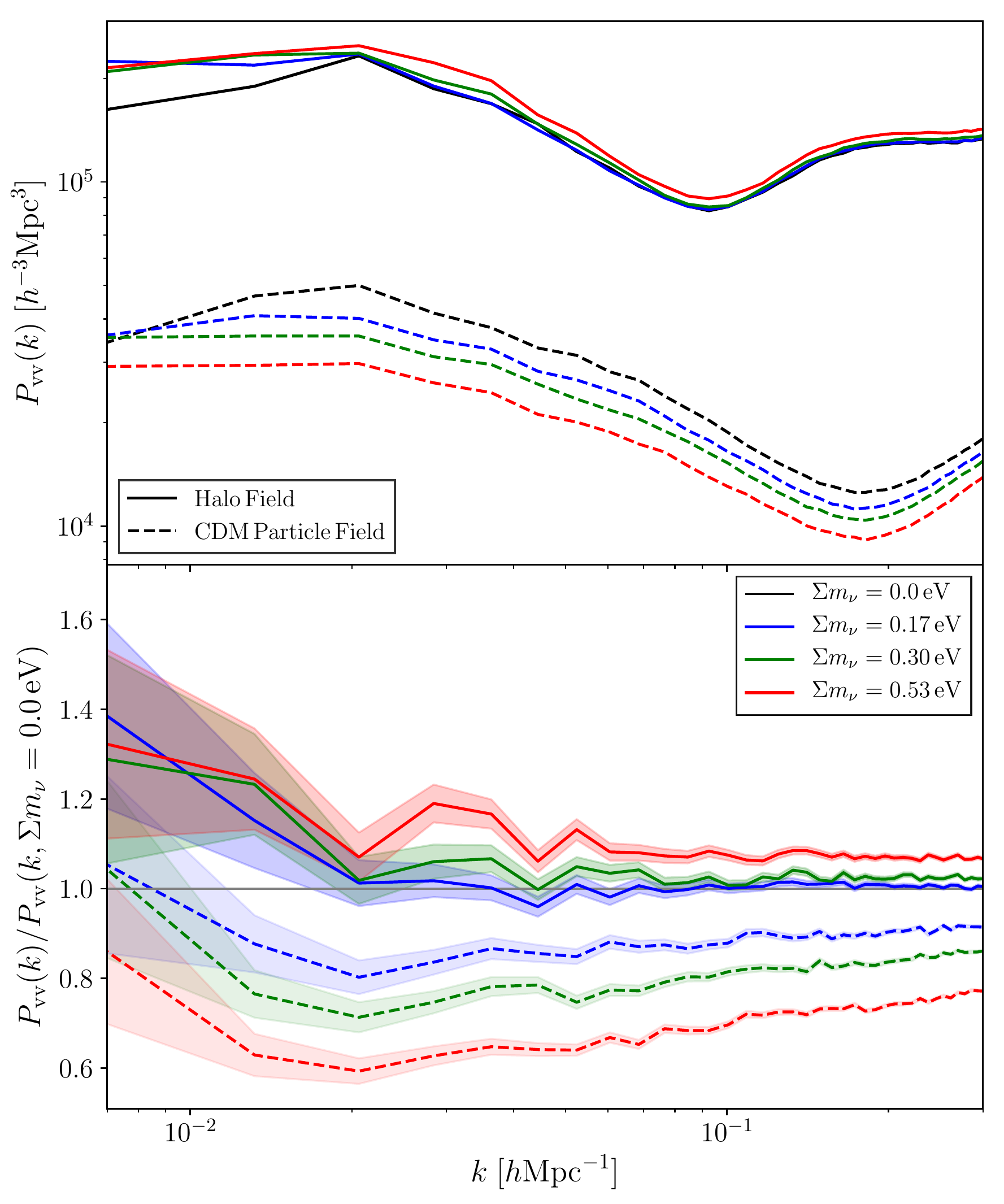}} \\
\subfigure[][]{%
\centering
\label{fig:PhhJia}%
\includegraphics[width=0.4\textwidth]{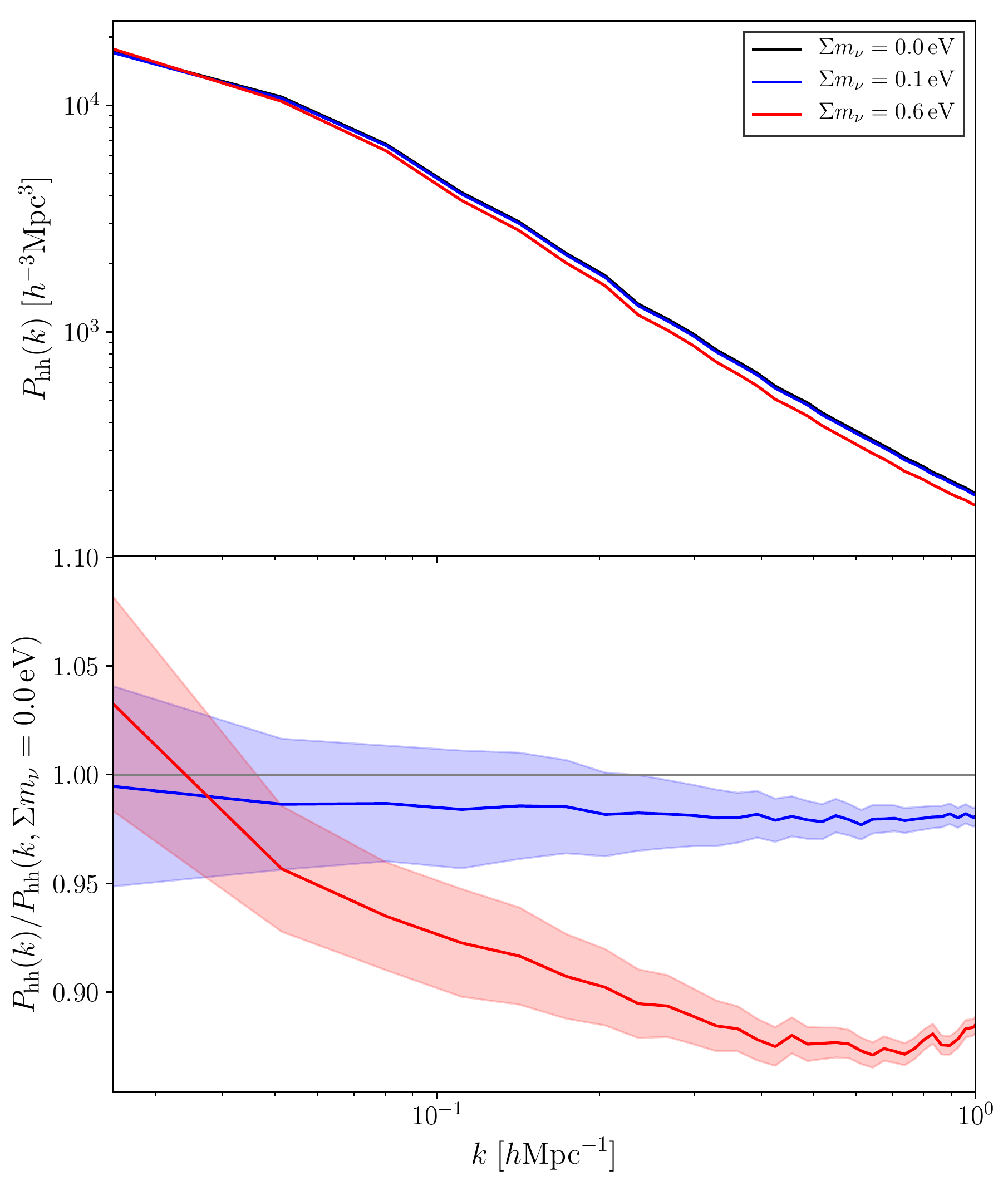}}%
\hspace{16pt}%
\subfigure[][]{%
\centering
\label{fig:PvvJia}%
\includegraphics[width=0.4\textwidth]{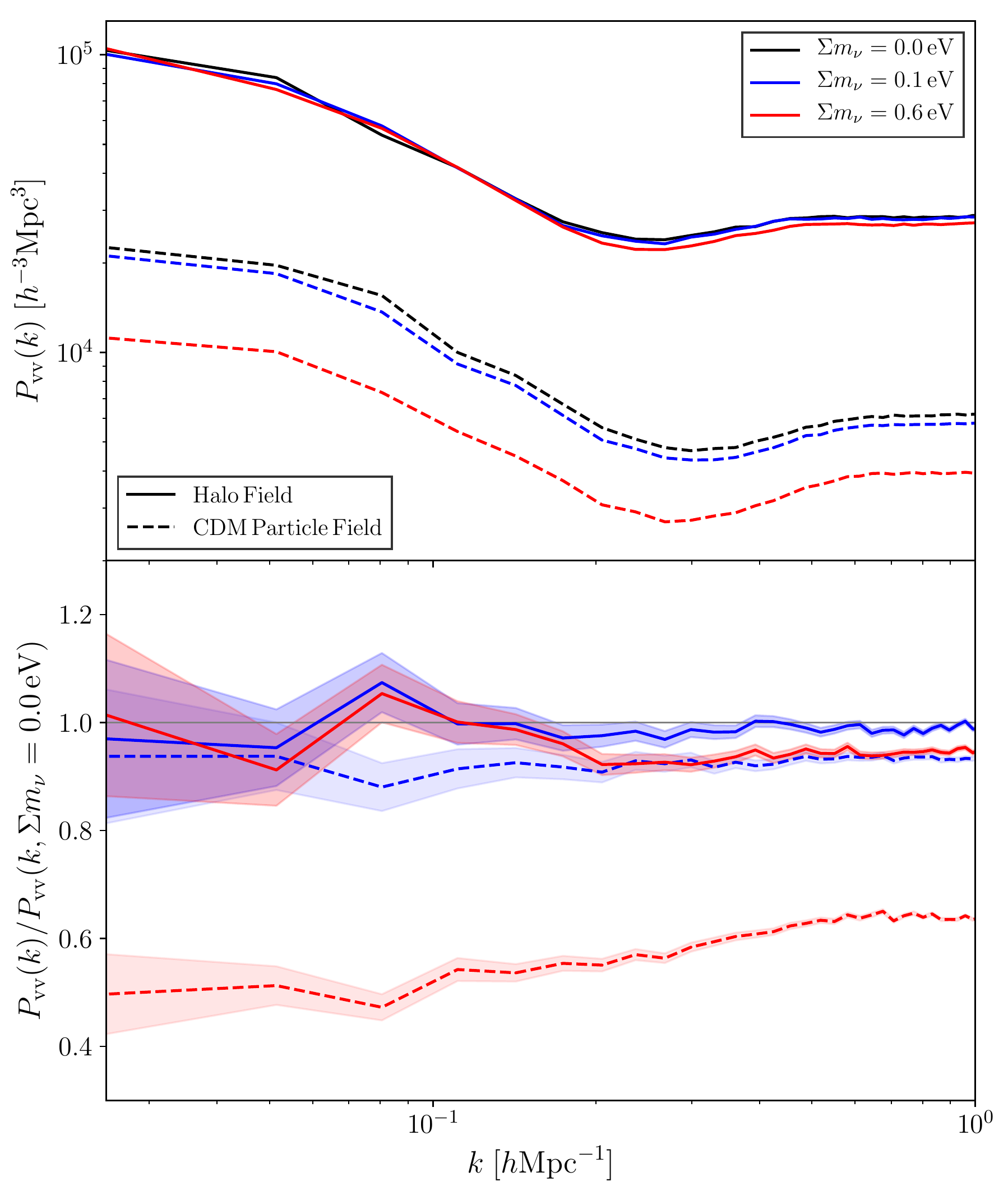}}%
\caption[]{
\subref{fig:PmmDEMNUni} The halo-halo power spectrum for the `low-res' simulation. Colors denote the sum of neutrino masses used in each simulation. The bottom panel shows the ratio between the different $\sum m_\nu$ cases and the massless case. Increasing $\sum m_\nu$ induces a biasing effect that \textit{boosts} the power spectrum. The power spectrum spans the scales accessible to the `low-res' simulation;
\subref{fig:PvvDEMNUni} Void-void power spectrum for the `low-res' simulation. Colors denote the sum of neutrino masses used in each simulation, dashed denotes voids traced by the CDM particle field, and solid denotes voids traced by the halo field. The bottom panel shows the ratio between the different $\sum m_\nu$ cases and the respective massless case. Increasing $\sum m_\nu$ \textit{boosts} the power spectrum for voids derived from the halos but \textit{damps} the power spectrum for voids derived from the particle distribution. The power spectrum spans the scales accessible to the `low-res' simulation;
\subref{fig:PhhJia} The halo-halo power spectrum for the `high-res' simulation. Colors denote the sum of neutrino masses used in each simulation. The bottom panel shows the ratio between the different $\sum m_\nu$ cases and the massless case. Increasing $\sum m_\nu$ \textit{damps} the power spectrum, in contrast to the effect on the `low-res' power spectrum. This is because the `high-res' simulation has a lower mass threshold ($M_\mathrm{min}=3 \times 10^{11}\,h^{-1}M_\odot$) than the `low-res' simulation ($M_\mathrm{min}=2.5 \times 10^{12}\,h^{-1}M_\odot$). The power spectrum spans the scales accessible to the `high-res' simulation, which are smaller than those for the `low-res' simulation since the `high-res' simulation has a smaller volume and larger resolution; and,
\subref{fig:PvvJia} The void-void power spectrum for the `high-res' simulation for voids derived from the halo distribution. Colors denote the sum of neutrino masses used in each simulation. The bottom panel shows the ratio between the different $\sum m_\nu$ cases and the massless case. Increasing $\sum m_\nu$ \textit{damps} the power spectrum, in contrast to the effect on the `low-res' power spectrum. We interpret this as due to the effective bias of the tracer population used to define voids (see Section \ref{sec:powerspectra}). The power spectrum spans the scales accessible to the `high-res' simulation.}%
\label{fig:PhhPvv}%
\end{figure*}

The void distribution is sensitive to $\sum m_\nu$ and how $\sum m_\nu$ impacts the underlying tracer distribution. Increasing $\sum m_\nu$ damps the CDM power spectrum, $P_\mathrm{cc}$, on small scales in the `low-res' simulation, as expected since neutrinos do not cluster on scales smaller than their free-streaming length \citep{Lesgourgues2006a}. As $\sum m_\nu$ increases, the effect becomes more significant.

\begin{table*}
	\centering
	\caption{{`Low-res'} average effective halo bias, $\bar{b}_\mathrm{h}$}
	\label{tab:bias_mnu_demnuni}
	\begin{tabular}{cccc} 
		\hline
		$\sum m_\nu = 0.0\,\mathrm{eV}$ & $\sum m_\nu = 0.17\,\mathrm{eV}$ & $\sum m_\nu = 0.30\,\mathrm{eV}$ & $\sum m_\nu = 0.53\,\mathrm{eV}$ \\
		\hline
		$1.003 \pm 0.002$ & $1.044 \pm 0.001$ & $1.083 \pm 0.001$ & $1.160 \pm 0.002$ \\
		\hline
	\end{tabular}
\end{table*}

\begin{table*}
	\centering
	\caption{{`High-res'} average effective halo bias, $\bar{b}_\mathrm{h}$}
	\label{tab:bias_mnu_jia}
	\begin{tabular}{ccc} 
		\hline
		$\sum m_\nu = 0.0\,\mathrm{eV}$ & $\sum m_\nu = 0.1\,\mathrm{eV}$ & $\sum m_\nu = 0.6\,\mathrm{eV}$ \\
		\hline
		$0.889 \pm 0.005$ & $0.898 \pm 0.005$ & $0.975 \pm 0.005$ \\
		\hline
	\end{tabular}
\end{table*}

The halo-halo power spectrum\footnote{All power spectra have a $k$ bin size $\Delta k\approx 0.008\,h\mathrm{Mpc}^{-1}$ unless otherwise noted and have uncertainties computed by \texttt{VIDE} and estimated from scatter in the bin average.}, $P_\mathrm{hh}$, for the `low-res' simulation shows an overall boost in power as $\sum m_\nu$ increases and biases the halo distribution (see Figure \autoref{fig:PmmDEMNUni}). Neutrinos reduce the growth of CDM perturbations. Therefore, at a fixed redshift, virialized halos have a smaller mass than in the massless neutrino case at the same redshift. The densest initial fluctuations in the matter density field will still form halos large enough to be detected in our simulations, but, depending on the value of $\sum m_\nu$, fluctuations with sufficiently low densities will no longer form halos with masses above the simulation mass threshold. Because only halos at the densest overdensities can be detected in simulations, halos at all scales are more highly correlated (with respect to the massless neutrino case), leading to a larger {effective} halo bias $b_\mathrm{h}$ {(see the average effective halo bias\footnote{{We compute the average effective halo bias by taking the average of $b_\mathrm{h} = \sqrt{P_{\rm hh}/P_{\rm cc}}$ for scales $k<0.1\,h\mathrm{Mpc}^{-1}$. We choose to cut at this scale to avoid the strongly nonlinear regime. All average effective halo biases have uncertainties computed from scatter in the bias values.}} for these simulations in \autoref{tab:bias_mnu_demnuni})}. The larger {effective} halo bias tends to compensate the suppression of the matter power spectrum due to free-streaming neutrinos, and the cumulative effect depends on $\sum m_\nu$. The halo power spectrum is given by:
\begin{align}
P_{\rm hh} = b_\mathrm{h}^2 P_{\rm cc},
\end{align}
where, in the presence of massive neutrinos, $b_\mathrm{h}$ is defined with respect to the cold dark matter density \citep{Castorina2013}. {See \citet{Hamaus2014a} for how void power spectra scale with $b_\mathrm{h}$. We denote $b_\mathrm{h}$ as the effective halo bias which can be assumed scale-independent at large scales when defined with respect to the cold dark matter field, but some scale-dependence can occur at smaller scales (see e.g. \citet{biasrev} and references therein, as well as \citet{Castorina2016} for the case of effective halo bias in the `low-res' simulation.) The `low-res' and `high-res' simulations nonetheless predominately cover scales where $b_\mathrm{h}$ is scale-independent.} The impact of the sum of neutrino masses on halo bias has been a topic of intense and ongoing study \citep[see e.g.][]{DeBernardis2008,Marulli2011,Villaescusa-Navarro2013,Castorina2013,Castorina2016,Biagetti2014,Loverde2014,Massara2015a,Petracca2015,Loverde2016,Desjacques2017,Raccanelli2017,Vagnozzi2018}. We note that a similar inversion in the effect of the sum of neutrino masses on the matter power spectrum and the halo power spectrum has been seen by \citet{Marulli2011} \citep[see also][]{Villaescusa-Navarro2013,Castorina2013}.

We find that increasing $\sum m_\nu$ boosts the correlation between voids derived from the halo distribution while it damps the correlation between voids derived from the CDM particle field for the DEMNUni simulation (see Figure \autoref{fig:PvvDEMNUni}).

To understand the effects of halo mass on the power spectra in the presence of neutrinos, we analyze the void distribution in the `high-res' simulations, which have a lower halo mass threshold. We plot the halo-halo power spectra and the void-void power spectra, as a function of $\sum m_\nu$ in Figure \autoref{fig:PhhJia} and Figure \autoref{fig:PvvJia}, respectively. The `high-res' simulations do not show the overall boost in the halo power for increasing $\sum m_\nu$ that we see in the `low-res' halo distribution. The void-void power spectra show a similar trend: for the `high-res' simulations, the power spectra of voids found in the halo distribution behave as the power spectra of voids derived from the CDM particle field, even if the differences due to neutrino effects are much less pronounced in the former than in the latter. In other words, the `high-res' void spectra \textit{do not show the same inversion} between the halo and CDM cases as that observed in the `low-res' simulations. {Impacts from the effective tracer bias cause this \textit{apparent} contradiction. The effective tracer bias for the `high-res' simulations is lower than that for the `low-res' simulations due to the lower minimum halo mass for the `high-res' simulations (compare the `low-res' average effective halo bias in \autoref{tab:bias_mnu_demnuni} to the `high-res' average effective halo bias in \autoref{tab:bias_mnu_jia}). We further discuss the physics behind this} in Section \ref{sec:discussion}.

\subsubsection{The Effects of Tracer Bias}
\label{sec:tracer_bias}

{Tracer bias impacts void statistics, and this manifests differently depending on the cosmological parameters probed. \citet{Pollina2015}, for example, found that coupled dark energy impacts voids found in the dark matter distribution but not voids found in the halos. They attribute this difference to the bias of the tracers used to find voids. In our case, we find that $\sum m_\nu$ has opposite impacts on voids found in the CDM and halo fields for the `low-res' simulation, and opposite impacts on voids found in the `high-res' halo field and voids found in the `low-res' halo field.}

\begin{table*}
	\centering
	\caption{{`Low-res'} average effective halo bias, $\bar{b}_\mathrm{h}$, for cut catalogs }
	\label{tab:bias_cut}
	\begin{tabular}{cccccc} 
		\hline
		 $ $ & $M \geq 2.5 \times 10^{12}\,h^{-1} M_\odot$ & $M \geq 5 \times 10^{12}\,h^{-1} M_\odot$ & $n_\mathrm{h}(M \geq 5 \times 10^{12}\,h^{-1} M_\odot)$ & $M \geq 1 \times 10^{14}\,h^{-1} M_\odot$ & $n_\mathrm{h}(M \geq 1 \times 10^{14}\,h^{-1} M_\odot)$ \\
		 $ $ & $\left(\mathrm{Original}\right)$ & $ $ & $\left(\mathrm{Random\,Subset}\right)$ & $ $ & $\left(\mathrm{Random\,Subset}\right)$ \\
		\hline
		$\sum m_\nu = 0.0\,\mathrm{eV}$ & $1.003 \pm 0.002$ & $1.111 \pm 0.002$ & $1.004 \pm 0.002$ & $2.24 \pm 0.01$ & $1.00 \pm 0.01$ \\
		$\sum m_\nu = 0.53\,\mathrm{eV}$ & $1.160 \pm 0.002$ & $1.316 \pm 0.001$ & $1.159 \pm 0.002$ & $2.80 \pm 0.01$ & $1.15 \pm 0.01$ \\
		\hline
	\end{tabular}
\end{table*}

While on the one hand neutrinos have a physical impact on the total number of voids (see Section \ref{sec:abundance}), on the other hand the number of voids directly maps to the void shot noise, which can be approximated at small scales as $1/n_{\rm v}$, where $n_{\rm v} = (\mathrm{Number\,of\,Voids})/\mathrm{Volume}$ is the void density.

To disentangle the impacts on the void power spectra of the void number and the effective halo bias as they change with $\sum m_\nu$, we remove shot noise and subsample the `low-res' simulation in two different manners:
\begin{enumerate}
\item we bias the halo distribution by making two mass cuts such that each of them contains only halos with $M \geq 5\times 10^{12} \,h^{-1} M_\odot$ or $M \geq 1\times 10^{14} \,h^{-1} M_\odot$;
\item we randomly subsample the halo distribution so that the number of halos matches that of the two subsamples defined in (i). In this way we produce sub-sets of halos with the same effective bias as the full halo distribution of the `low-res' simulations, but with the same halo number density as the highly biased subsamples in (i) (see e.g. \autoref{fig:PvvJia_cuts} in Appendix \ref{sec:vol_res} for a similar application to the `high-res' simulations).
\end{enumerate}
{The average effective bias for each of these cuts is shown in \autoref{tab:bias_cut} for the `low-res' massless and $\sum m_\nu = 0.53\,\mathrm{eV}$ neutrino simulations. The randomly subsampled halos have the same effective bias as the original simulation, and the mass cut halos have higher biases.}

To remove the effects of void number density we model the shot noise for the void-void power spectrum as scale-dependent following the prescription by {\citet{Hamaus2010} and} \citet{Hamaus2014a}, which is well approximated by $1/n_{\rm v}$ for small scales:
\begin{align}
\mathcal{E}_{\rm vv}(k)=P_{\rm vv}-\frac{P_{\rm vc}^2}{P_{\rm cc}},
\end{align}
where $P_{\rm vv}$ is the void-void power spectrum and $P_{\rm vc}$ is the void-CDM cross-correlation power spectrum. Thus, we can write the void power spectrum with shot noise removed as
\begin{align}
P_{\rm vv,\mathrm{no\,shot}}(k) = \frac{P_{\rm vc}^2}{P_{\rm cc}}.
\end{align}

\begin{figure}
\begin{center}
\includegraphics[width=0.5\textwidth]{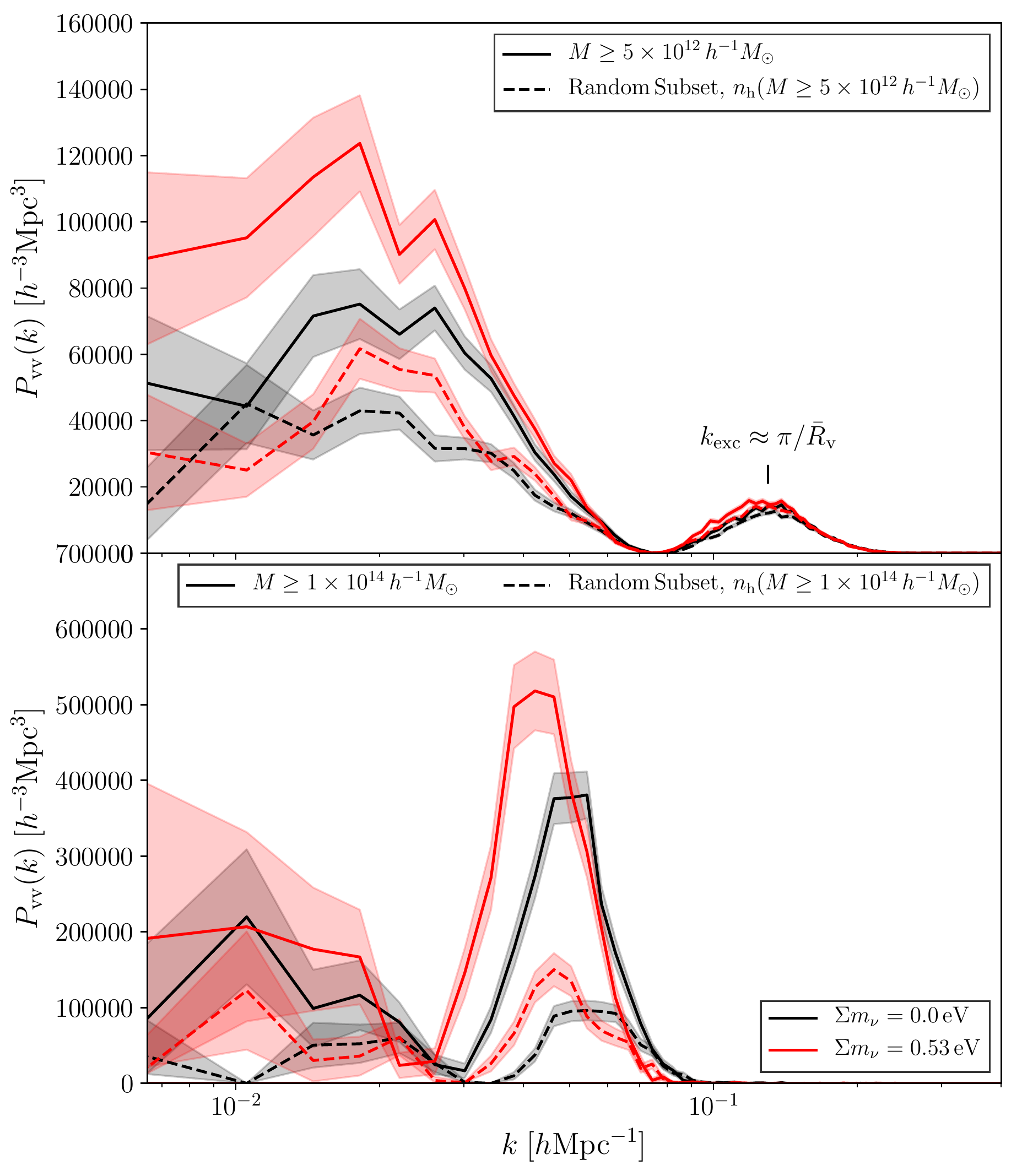}
\caption{The void-void power spectrum with shot noise removed for the `low-res' simulation for voids derived from the halo distribution. Removing shot noise removes the effects of void number density. Colors denote the sum of neutrino masses used in each simulation. The top panel corresponds to voids found in the less highly biased tracer field, while the bottom panel corresponds to voids found in the highly biased tracer field. Dashed lines correspond to randomly subsampling the original halo catalog so its number density matches that of the mass thresholded catalog, removing the effects of tracer density. The impact of $\sum m_\nu$ on void clustering depends on the effective halo bias.}
\label{fig:PvvDEMNUni_noshot}
\end{center}
\end{figure}

The sum of neutrino masses affects the amplitude and phase of the void-void power spectrum. In \autoref{fig:PvvDEMNUni_noshot} we plot the void power spectra (with shot noise removed) for the two highly biased catalogs of (i), and compare them with the void spectra of the corresponding randomly subsampled catalogs of (ii) (see \autoref{fig:PvvJia_biascuts} in Appendix \ref{sec:vol_res} for analogous void power spectra including shot noise for multiple halo mass thresholds from the `high-res' simulation).
{At scales larger than the void exclusion scale}, the void power spectrum tracks the tracer power spectrum {\citep{Chan2014,Clampitt2016}}: the power at large scales for the voids traced by halos with a higher mass threshold is larger, as expected for a more biased sample. Nonetheless, the large scale power is of the same order of magnitude for both the mass thresholds at large scales (compare top and bottom panels of \autoref{fig:PvvDEMNUni_noshot}). For the highly biased tracers ($M \geq 1 \times 10^{14}\,h^{-1} M_\odot$, bottom panel), the power at large scales is dominated by uncertainties because there are less small voids that correlate at large scales. For the less highly biased tracers ($M \geq 5 \times 10^{12}\,h^{-1} M_\odot$, top panel) there is a discernible difference for the two neutrino masses at large scales because there is a large number of small voids traced by smaller halos, improving the uncertainties.

\begin{figure}
\begin{center}
\includegraphics[width=0.5\textwidth]{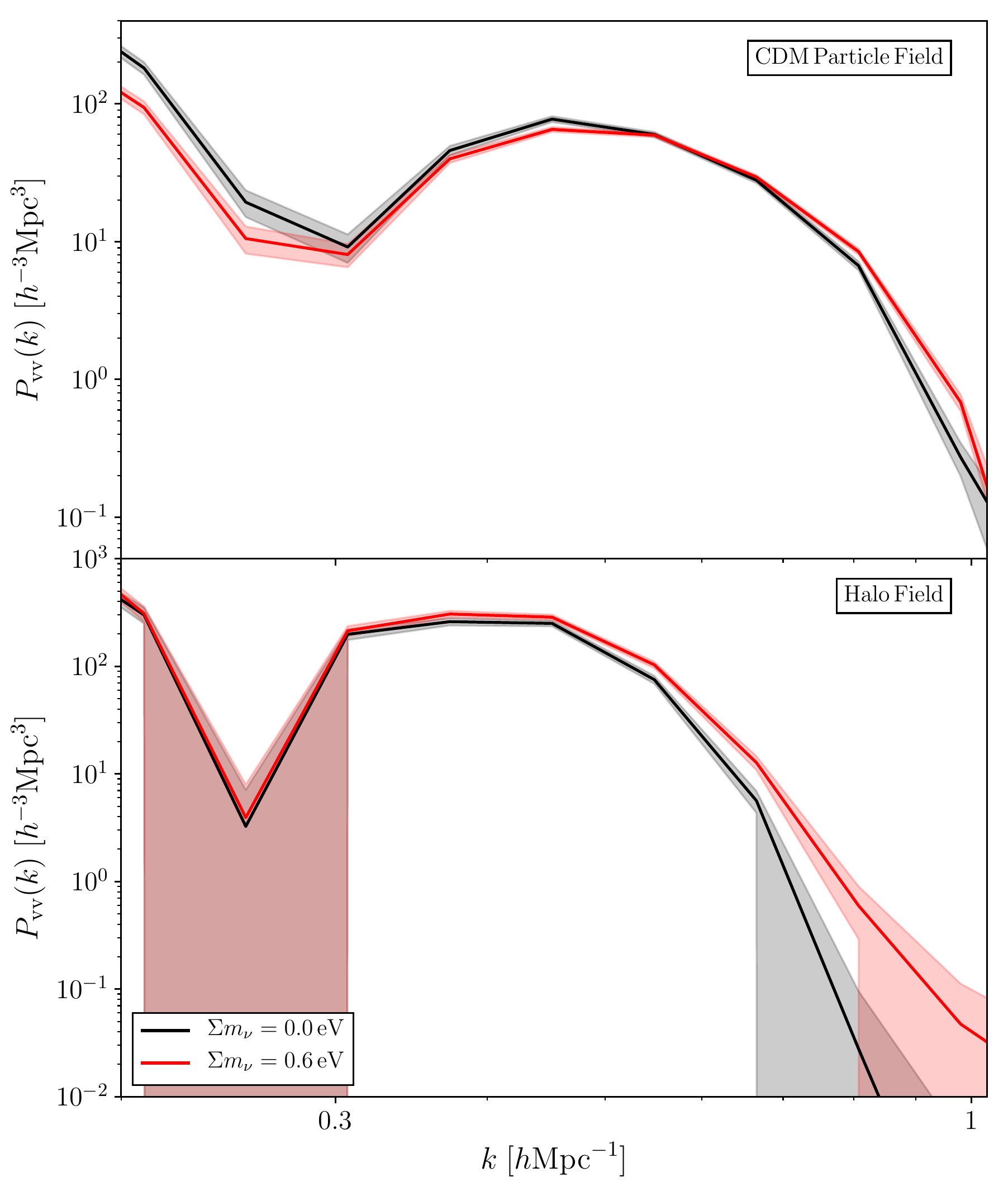}
\caption{The void-void power spectrum near the exclusion scale with shot noise removed for the `high-res' simulation. Colors denote the sum of neutrino masses used in each simulation. The top panel corresponds to voids found in the CDM particle field, while the bottom panel corresponds to voids found in the halo field. The low mass threshold $M \geq 3 \times 10^{11}\,h^{-1}M_\odot$ for the `high-res' simulation causes voids traced by the halos to behave similar to voids traced by the CDM particles for small scales.}
\label{fig:PvvMassiveNuS_noshot}
\end{center}
\end{figure}

The power at small scales dramatically increases with $\sum m_\nu$ when increasing the effective halo bias (compare top and bottom panels of \autoref{fig:PvvDEMNUni_noshot}). The small voids that remain when increasing $\sum m_\nu$ have highly biased halos forming their walls. These highly biased halos sit near overdensities, forming a concentrated cosmic web with voids that are, thus, tightly packed, boosting their correlation. The minimum at scales just larger than $k=10^{-1}\,h\mathrm{Mpc}^{-1}$ corresponds to the scale at which voids are uncorrelated \citep[see e.g.][]{Hamaus2014a}. The scale of the local maximum to the right of this minimum corresponds to the void exclusion scale, $k_\mathrm{exc}\approx \pi/\bar{R}_\mathrm{v}$, where $\bar{R}_\mathrm{v}$ is the average void radius. This is the smallest scale at which voids with radius $\bar{R}_\mathrm{v}$ do not overlap \citep{Hamaus2014a}.

Increasing $\sum m_\nu$ shifts the power from small scales to large scales for the `low-res' voids found in the halo distribution. $\sum m_\nu$ may create a scale-dependent bias in voids, but this effect must be more thoroughly investigated to determine if the scale dependence is due to neutrino properties, non-linearities, or other effects. Increasing the effective halo bias increases the scale-dependent impact $\sum m_\nu$ has on the void power spectra. This is seen most clearly near the void exclusion scale. This shift in power from small voids to large voids is consistent with $\sum m_\nu$ decreasing the number of small voids and increasing the number of large voids for voids derived from the halo distribution in the `low-res' simulations, thus causing the average void radius to increase and $k_\mathrm{exc}$ to decrease (see Section \ref{sec:abundance}).

\begin{figure}
\begin{center}
\includegraphics[width=0.5\textwidth]{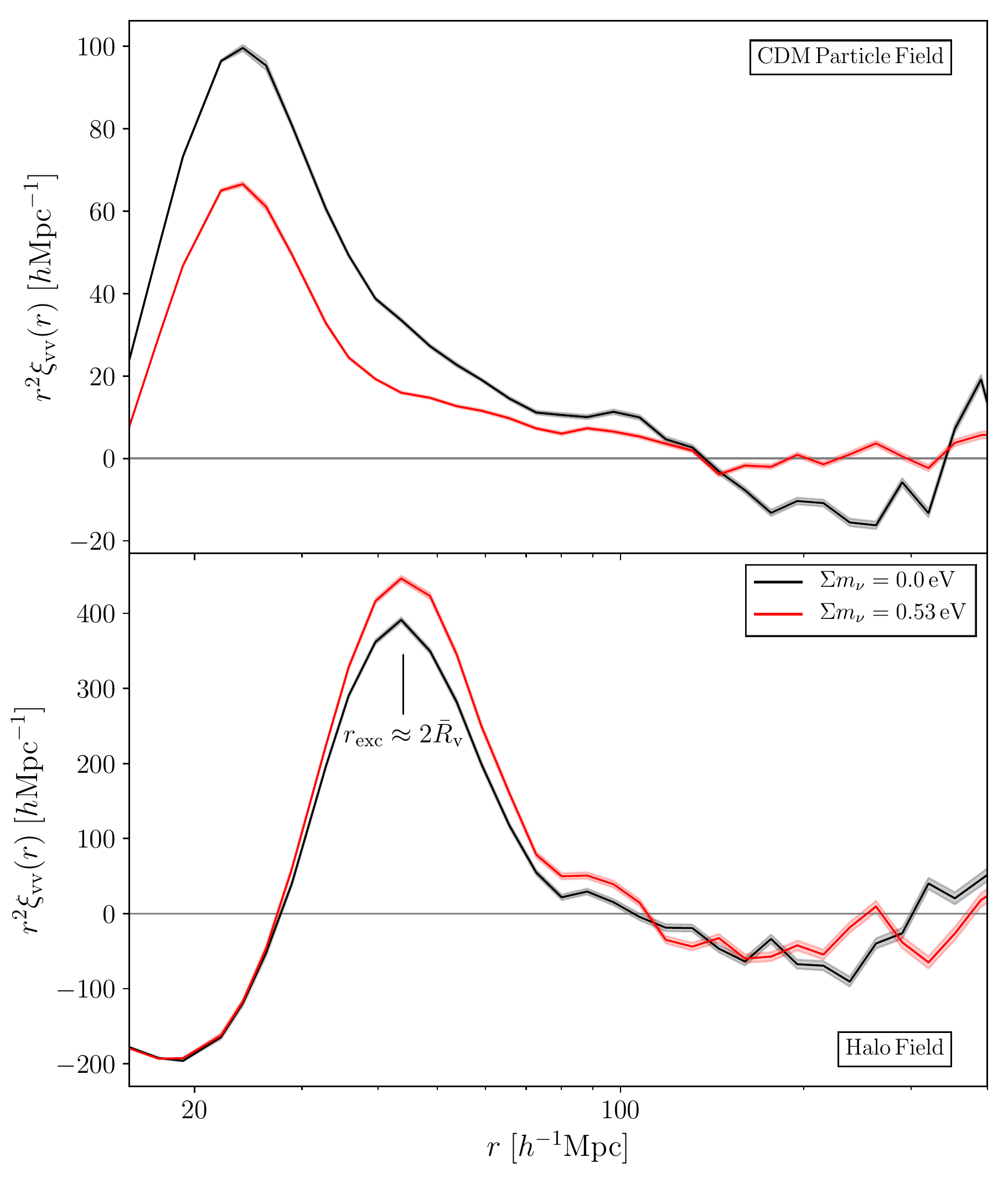}
\caption{The void auto-correlation function for `low-res' voids, including uncertainties. We scale the correlation functions by $r^2$ to emphasize the effects at large $r$. Colors denote the sum of neutrino masses used in each simulation. The top panel corresponds to voids found in the CDM particle field, while the bottom panel corresponds to voids found in the halo field. Increasing $\sum m_\nu$ diminishes void clustering for voids traced by CDM particles while it enhances void clustering for voids traced by halos. All correlation functions are cut at 2 times the mean particle separation in the simulation and where scales are so large that noise dominates. Voids traced by the CDM particles are so small that the correlation function does not become negative for scales larger than 2 times the particle separation due to the simulation resolution.}
\label{fig:XivvDEMNUni_CDM}
\end{center}
\end{figure}

\begin{figure}
\begin{center}
\includegraphics[width=0.5\textwidth]{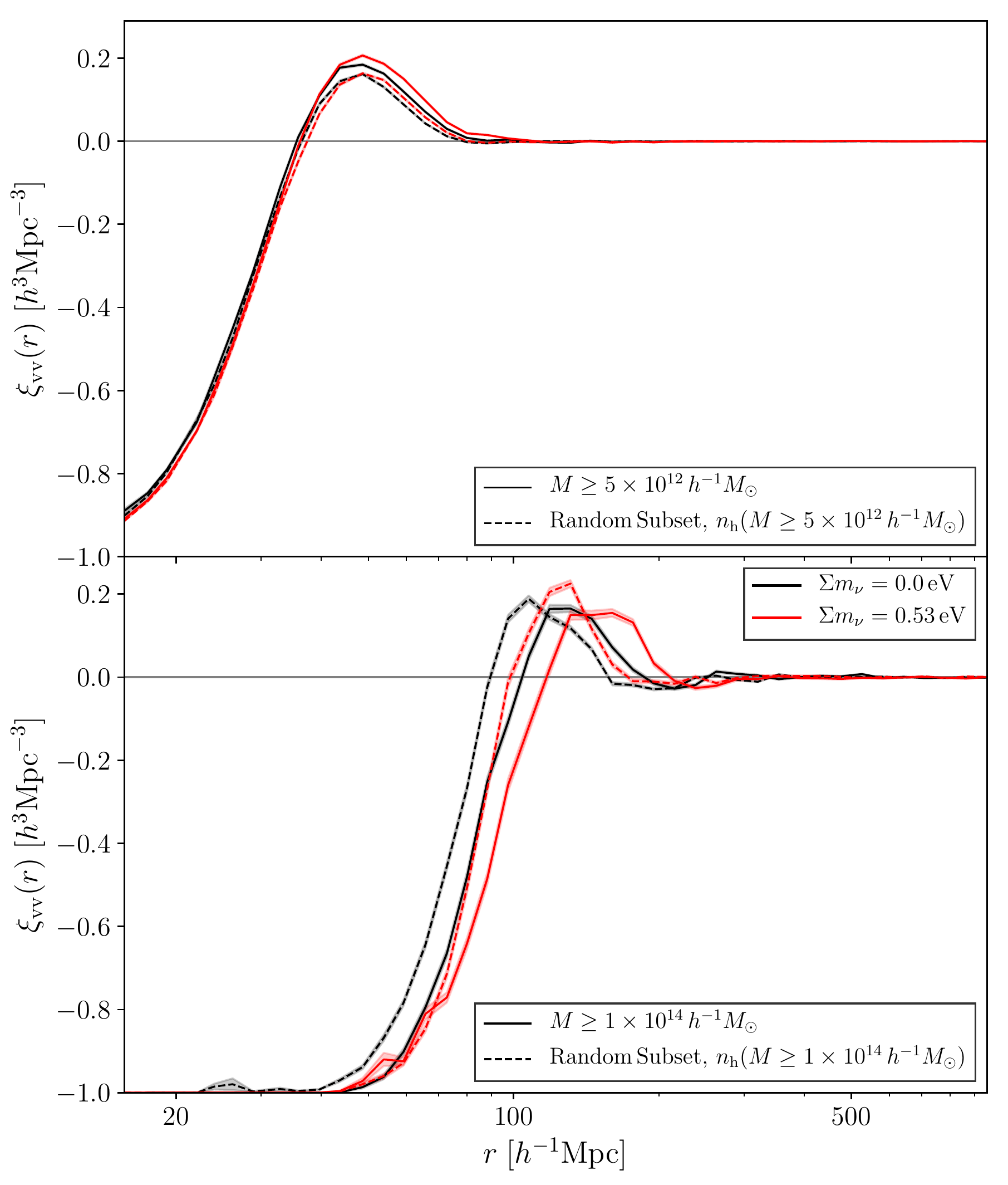}
\caption{The void auto-correlation function for the `low-res' simulation for voids derived from the halo distribution, including uncertainties. Colors denote the sum of neutrino masses used in each simulation. The top panel corresponds to voids found in the less highly biased tracer field, while the bottom panel corresponds to voids found in the highly biased tracer field. Increasing $\sum m_\nu$ shifts the correlation peak to larger scales and boosts the correlation. Increasing the effective halo bias amplifies the effect of $\sum m_\nu$ on void clustering.}
\label{fig:XivvDEMNUni}
\end{center}
\end{figure}

On the other hand we find that, for the `high-res' simulations, increasing $\sum m_\nu$ shifts the power in the void power spectra (with shot noise removed) from large to small scales for both the CDM voids and halo voids (see near the exclusion scale $k\approx 0.5\,h\mathrm{Mpc}^{-1}$ in \autoref{fig:PvvMassiveNuS_noshot}, which has bin size $\Delta \log k\approx 0.08\,h\mathrm{Mpc}^{-1}$). This is in contrast to the shift in power from small to large scales seen for the `low-res' simulation in \autoref{fig:PvvDEMNUni_noshot}. We note that the `low-res' void power spectra (with shot noise removed) for CDM voids is consistent with that of the `high-res' simulation.

Tracer bias influences how different kinds of voids respond to $\sum m_\nu$: a low mass threshold, and so a low effective tracer bias, does not produce an inversion between the CDM case and halo case for the void abundance and power spectra. We have verified that sampling the `high-res' halo distribution so it has the same minimum halo mass as the `low-res' simulation, $M \geq 2.5\times 10^{12} \,h^{-1} M_\odot$ and thus increasing the effective tracer bias, leads to the inverted behavior between the biased halo case and the CDM case for the abundances, total number of voids (see \autoref{fig:void_number}), and the power spectra, like seen for the `low-res' simulation. The exclusion scale in the biased `high-res' distribution also shifts from small scales to match the `low-res' exclusion scale.

The correlation functions are a useful tool to view the $\sum m_\nu$ inversion effects in real space. In \autoref{fig:XivvDEMNUni_CDM} we plot the void auto-correlation function\footnote{All correlation functions are computed by \texttt{VIDE} via an inverse Fourier transform of the power spectra, have an $r$ bin size $\Delta \log r\approx 0.04\,h^{-1}\mathrm{Mpc}$, and have uncertainties computed by \texttt{VIDE} and estimated from scatter in the bin average.} for voids derived from the CDM particle field and the halo field. $\xi_{\rm vv}$ peaks at the void exclusion scale $2\bar{R}_{\rm v}$ because this is the average distance at which voids are most tightly packed, i.e. the walls of neighboring spherical voids with a radius equal to the average void radius meet. $\xi_{\rm vv}$ decreases for smaller scales, i.e. scales smaller than $2\bar{R}_{\rm v}$, since voids do not overlap. As explained in \citet{Massara2015}, this decline is gradual because voids are not perfect spheres and they have different sizes. For scales larger than the exclusion scale, voids do not cluster as much and so $\xi_{\rm vv}$ falls. We note that the void auto-correlation function becomes negative at scales larger than the Baryon Acoustic Oscillations (BAO) before approaching zero since voids trace the matter distribution at large scales {\citep{Chan2014,Clampitt2016}}. Voids are not likely to be separated by this distance.

Increasing $\sum m_\nu$ suppresses void clustering for the CDM case at scales smaller than the BAO peak position, and reduces the anticorrelation at large scales since there are more voids spread throughout the field. Voids derived from the halos cluster more near the exclusion scale, showing opposite behavior to the CDM case just like the power spectra.

In the upper panel of \autoref{fig:XivvDEMNUni} we compare, for two different values of $\sum m_\nu$, voids derived from the less highly biased halo catalog defined in (i) to the corresponding catalog, defined in (ii), derived from the original halo catalog with the same halo density for two different $\sum m_\nu$. Increasing $\sum m_\nu$ boosts the correlation of voids derived from the biased halo sample, analogous to the effect on halos with large effective bias. Increasing the neutrino mass reduces the number of small voids traced by halos in the field, so the remaining voids are more highly correlated, resulting in a higher correlation peak. Since there are less small voids and more large voids, there is more void clustering for scales larger than the exclusion scale.

$\sum m_\nu$'s impacts on the amplitude and scale are most prominent for voids traced by highly biased tracers. In the lower panel of \autoref{fig:XivvDEMNUni} we show the void auto-correlation function for voids derived from the highly biased halo sample and the original catalog with the same halo density. Decreasing the tracer density and increasing the effective halo bias both shift the average void radius to larger scales, causing the correlation function to peak at larger scales (compare upper and lower panels). For the voids traced by the less dense and highly biased halos, increasing $\sum m_\nu$ strongly shifts the entire correlation function to larger scales, similarly to the impact on the power spectra in \autoref{fig:PvvDEMNUni_noshot}.

The impact of $\sum m_\nu$ on the correlation functions is not simply explained by the effects of void abundance. In the upper panel of \autoref{fig:XivvDEMNUni} we see that increasing $\sum m_\nu$ boosts the correlation for the voids traced by the less highly biased halos without significantly changing the peak location relative to the massless case. On the contrary, for the highly biased case in the lower panel of \autoref{fig:XivvDEMNUni}, the amplitude at the correlation peak does not change between the massless and $\sum m_\nu = 0.53\,\mathrm{eV}$ cases.  If void abundance solely drove $\sum m_\nu$'s impacts on the correlation functions, the correlation peak's amplitude would decrease as the average void radius increases \citep[see e.g. the void auto-correlation functions in][for different void sizes]{Massara2015}. Neutrinos impact the clustering of voids -- $\sum m_\nu$ influences void bias \citep[see e.g.][]{voidbias}. We further explore the impact of $\sum m_\nu$ in our upcoming paper.

{ How distinct is this fingerprint? \citet{Massara2015} investigated the degeneracy between $\sum m_\nu$ and $\sigma_8$ for voids found in the dark matter field. They found that altering $\sigma_8$ cannot reproduce the effects of $\sum m_\nu$ on void properties like number density, density profiles, and velocity profiles. For voids found in the `low-res' halo field, we find that the dominant impacts of $\sum m_\nu$ on void clustering occur for scales $k \lesssim 0.1\,h\mathrm{Mpc}^{-1}$ (see \autoref{fig:PvvDEMNUni_noshot}). These are exactly the scales for which the effects of $\sum m_\nu$ and $\sigma_8$ become distinct in the `low-res' halo power spectra \citep[see Figure 8 in][]{Castorina2016}.

Further, the response of void clustering to $\sum m_\nu$ changes sign as a function of the effective halo bias, a trend uncommon for cosmological parameters like $\sigma_8$. These trends and previous studies suggest that the impacts we see on voids from $\sum m_\nu$ are distinct from those of $\sigma_8$.

Finally, as extended 3-dimensional objects, voids must be defined by 4 (non-planar) points. Thus, voids contain information about the 3- and 4-point clustering of the tracers, and as such provide \textit{information beyond the tracer 2-point clustering.} The void exclusion scale is a manifestation of this and shifts in response to $\sum m_\nu$. Halos do not have an equivalent feature or response. For these reasons, $\sum m_\nu$ leaves distinct fingerprints on voids.}

\section{Discussion}
\label{sec:discussion}
Our work indicates that voids respond to $\sum m_\nu$ in two distinct manners, determined by if they are derived from the halo distribution or the cold dark matter particle field. Both the halo and CDM distributions should be utilized to properly study voids and the impact neutrinos have on them. For forecasting constraints on $\sum m_\nu$, the void catalog ideally should be built from the survey mock or HOD populated simulation rather than the CDM distribution.

Increasing $\sum m_\nu$ slows down the growth of the CDM perturbations, reducing the CDM overdensities present today. Since the evolution of the overdensities has slowed, fewer mergers of the small overdensities have occurred, resulting in a larger number of small CDM overdensities and fewer large CDM overdensities relative to the massless neutrino case. The numerous smaller CDM overdensities yield smaller voids since the small overdensities fragment what would be large voids. Hence, increasing $\sum m_\nu$ increases the number of small voids and decreases the number of large voids derived from the CDM particle field. Since there are more small overdensities in the field as $\sum m_\nu$ increases, voids become less biased near the correlation peak since they are not as localized and less antibiased for scales larger than the BAO peak position, as it is more likely to find voids separated by larger distances.

We note that our void finding procedure in the CDM case only uses CDM particles and does not include the neutrino particles. A different approach is to locate voids in the total matter field, such as in the work of \citet{Banerjee2016} that included neutrino particles and CDM particles. In our work, we have established that the inversion is unique to voids derived from halos because the effective halo bias drives the inversion. Therefore, our results are particularly relevant to interpreting void observations.

For the halo case, increasing $\sum m_\nu$ makes halos less massive, leaving only the halos that sit at large density perturbations detectable in our simulations. Thus, these halos are more highly correlated and we see a bias effect in the halo-halo power spectra. For the \texttt{DEMNUni} simulation, only massive halos remain due to the limited mass resolution of the simulations, so there are no longer small halos that could segment a larger void into separate voids. For this reason and since larger voids are defined by larger overdensities, increasing $\sum m_\nu$ increases the number of large voids derived from the halo catalog and decreases the number of small voids.

The high resolution of the `high-res' simulation produces a lower minimum halo mass and, thus, halos that are less biased tracers of the CDM particle field than the `low-res' simulation. The `high-res' simulation can identify halos at smaller CDM overdensities than the `low-res' simulation, and, consequently, these halos have masses and an effective bias lower than the `low-res' mass resolution. However, the `high-res' simulation has a finite resolution and cannot identify halos at the smallest CDM overdensities, so its halo catalog is still biased (even if its effective bias is smaller than the `low-res' halo catalogs), and its halos have a higher correlation than the CDM overdensities.

Since increasing $\sum m_\nu$ leads to more small CDM overdensities and the `high-res' simulation has a low effective halo bias, `high-res' halos trace these small CDM overdensities more than the `low-res' halos. Halos in the `high-res' simulation are less biased tracers of the matter density field; therefore, the increased correlation due to the halo's effective bias from the simulation resolution and $\sum m_\nu$ is not substantial enough to overpower the damping effects from the neutrino free-streaming. Thus, the `high-res' void power spectra for voids found in the CDM field and for voids found in the halo field damp as $\sum m_\nu$ increases.

\section{Conclusions \& future prospects}
\label{sec:conclusions}
We have explored the impact of the sum of neutrino masses $\sum m_\nu$ on void properties with the N-body simulations \texttt{DEMNUni} and \texttt{MassiveNuS}. For the first time we have shown that:
\begin{enumerate}
\item the effect $\sum m_\nu$ has on void properties depends on the type of tracer the void catalog was built from,
\item using voids only derived from the cold dark matter particle field to study neutrinos, as has been assumed in the literature, is not sufficient to capture the effects of neutrinos on voids. Voids are not always smaller and denser in the presence of neutrinos, and tracer properties can actually lead to larger voids, a smaller number of voids, and enhanced void clustering,
\item the impact of $\sum m_\nu$ on the void abundance and void-void power spectrum for the \texttt{DEMNUni} {(`low-res')} void catalog derived from the halo distribution is opposite to that for the void catalog derived from the CDM particle field. For voids derived from the cold dark matter field, increasing $\sum m_\nu$ increases the number of small voids, decreases the number of large voids, and damps the void-void power spectrum. The opposite is true for voids derived from the biased halo distribution due to the effects of the effective halo bias,
\item the effective halo bias influences how $\sum m_\nu$ affects voids -- this will have interesting impacts on future surveys aiming to constrain the sum of neutrino masses, and
\item void power spectra and auto-correlation functions are powerful tools for distinguishing neutrino masses. Neutrinos leave a distinct fingerprint on voids, which can potentially help break the degeneracy between cosmological parameters in halo measurements. We plan to thoroughly explore breaking degeneracies, such as $\sigma_8$, in upcoming work.
\end{enumerate}

By comparing observations of the number of voids, void abundance, and void clustering to $\Lambda\mathrm{CDM}$ simulations with volume and resolution matching the survey volume and galaxy number density, surveys have a new avenue to place constraints on $\sum m_\nu$. { It is important to note, though, that for a fixed volume, substantially low tracer densities produce large measurement uncertainties due to a small number of voids. Thus, surveys with low tracer densities in combination with smaller volumes relative to those shown in this work may not be able to statistically distinguish the impacts neutrinos have on voids.} However, upcoming surveys like PFS, DESI, { and} Euclid have halo densities {$n_{\rm h}$ of $\approx 6 \times 10^{-4}\,h^3\mathrm{Mpc}^{-3}$ \citep{PFS}, $7 \times 10^{-4}\,h^3\mathrm{Mpc}^{-3}$ \citep{DESI}, and $2 \times 10^{-3}\,h^3\mathrm{Mpc}^{-3}$ \citep{WFIRST}, respectively, for $z\approx 1$} comparable to that of the \texttt{DEMNUni} {(`low-res')} simulation { with $n_{\rm h}\approx 1 \times 10^{-3}\,h^3\mathrm{Mpc}^{-3}$ at $z=1.05$.} {Denser surveys like WFIRST with $n_{\rm h}\approx 9 \times 10^{-3}\,h^3\mathrm{Mpc}^{-3}$ \citep{WFIRST} at the same redshift} can even exceed the \texttt{DEMNUni} {(`low-res')} simulation's density{. Thanks to their high tracer densities and large volumes, these surveys will be capable of measuring the impact $\sum m_\nu$ has on voids}. For these upcoming observations, simulations such as \texttt{DEMNUni} and \texttt{MassiveNuS} are the best tools for evaluating the impact of neutrinos on the observed voids. In the final stages reliable mocks will also be necessary to correctly evaluate the mask and survey boundary effects.

The opposite behavior of the \texttt{DEMNUni} {(`low-res')} and \texttt{MassiveNuS} {(`high-res')} simulations to $\sum m_\nu$ indicates there exists a threshold effective halo bias for which the void power spectra, correlation functions, and abundances for voids derived from the halo distribution will be less sensitive to $\sum m_\nu$. It would be interesting to compare surveys with effective halo biases above and below the threshold at which $\sum m_\nu$ induces the inversion effect in the void abundances, number, power spectra, and correlation functions, since lower densities increase the minimum halo mass, and so the effective halo bias, of the survey. In this sense one could imagine an extraordinarily dense low-$z$ survey to be particularly interesting. Within the same survey, it will be interesting to compare void properties for tracers with different luminosity or mass thresholds, i.e. with different biases. The use of multi-tracer techniques is another promising tool for constraining $\sum m_\nu$ and its impact on voids. Utilizing the redshift dependence of these effects and redshift coverage of these surveys could further yield unique constraints on neutrino properties. We explore this interdependence in our upcoming paper.

\section*{Acknowledgements}

{We thank the anonymous referee for their helpful comments.} We thank F. Villaescusa-Navarro and J. Bel for useful discussions and comments on the manuscript. CDK is supported by the National Science Foundation Graduate Research Fellowship under Grant DGE 1656466. AP and EM are supported by NASA grant 15-WFIRST15-0008 to the WFIRST Science Investigation Team ``Cosmology with the High Latitude Survey". The DEMNUni-I simulations were carried out at the Tier-0 IBM BG/Q machine, Fermi, of the Centro Interuniversitario del Nord-Est per il Calcolo Elettronico (CINECA, Bologna, Italy), via the ve million cpu-hrs budget provided by the Italian SuperComputing Resource Allocation (ISCRA) to the class-A proposal entitled ``The Dark Energy and Massive-Neutrino Universe". C.C. acknowledges financial support from the European Research Council through the Darklight Advanced Research Grant (n. 291521). JL is supported by an NSF Astronomy and Astrophysics Postdoctoral Fellowship under award AST-1602663. This work used the Extreme Science and Engineering Discovery Environment (XSEDE), which is supported by NSF grant ACI-1053575. The \texttt{MassiveNuS} simulations are publicly available at \url{http://columbialensing.org} through the Skies \& Universes Project. This work has been done within the Labex ILP (reference ANR-10-LABX-63) part of the Idex SUPER, and received financial state aid managed by the Agence Nationale de la Recherche, as part of the programme Investissements d'avenir under the reference ANR-11-IDEX0004-02. The Flatiron Institute is supported by the Simons Foundation.

\appendix
\section{Simulation and Void Finder Details}
\label{sec:sim_details}

\subsection{The \texttt{DEMNUni} simulation suite}
\label{sec:nbody}

The \texttt{DEMNUni} simulations have been performed using the tree particle mesh-smoothed particle hydrodynamics (TreePM-SPH) code
GADGET-3 \cite{Springel2001}, specifically modified by~\cite{Viel2010} to account for the presence of massive neutrinos. They are characterized by a softening length $\varepsilon=20 \mathrm{kpc}$, start at $z_{\rm in}=99$, and are performed in a cubic box of side $L = 2000\,h^{-1}\mathrm{Mpc}$, containing  $N_{\rm p} = 2048^3$ CDM particles, and an equal number of neutrino particles when $\sum m_\nu \neq 0 $ eV. These features make the \texttt{DEMNUni} set suitable for the analysis of different cosmological probes, from galaxy-clustering, to weak-lensing, to CMB secondary anisotropies.

Halos and sub-halo catalogs have been produced for each of the 62 simulation particle snapshots, via the friends-of-friends (FoF) and SUBFIND algorithms included in Gadget III \cite{Springel2001,Dolag2010}. The linking length was set to be $1/5$ of the mean inter-particle distance \citep{Davis1985} and the minimum number of particles to identify a parent halo was set to 32, thus fixing the minimum halo mass to $M_{\rm FoF}\simeq 2.5\times 10^{12}$ \hmone $M_\odot$.

\subsection{The \texttt{MassiveNuS} simulation suite}

The \texttt{MassiveNuS} simulations consists a large suite of 101 N-body simulations, with three varying parameters $\sum m_\nu$, $A_s$, and $\Omega_m$. In order to avoid shot noise and high computational costs typically associated with particle neutrino simulations, \texttt{MassiveNuS} adopts a linear response algorithm~\citep{Ali-Haimoud2013}, where neutrinos are described using linear perturbation theory and their clustering is sourced by the full non-linear matter density. This method has been tested robustly against CDM particle simulations and agreements are found to be within 0.2\% for $\sum m_\nu \leq 0.6$ eV.

The simulations use the public code \texttt{Gadget-2}, patched with the public code \texttt{kspace-neutrinos} to include neutrinos\footnote{The code also has the flexibility to include neutrinos as particles at low redshifts, to capture neutrino self-clustering. The latest version may be found here: \url{https://github.com/ sbird/kspace-neutrinos}}. The \texttt{MassiveNuS} halo catalogues are computed using the public halo finder code \texttt{Rockstar}\footnote{\url{https://bitbucket.org/gfcstanford/rockstar}}~\citep{Behroozi2013}, also a friends-of-friends-based algorithm.

\subsection{Void finder}
\label{sec:vide}
\texttt{VIDE} performs a Voronoi tessellation of the tracer field, creating basins around local minima in the density field. It then relies on the Watershed transform (Platen et al) to merge basins and construct a hierarchy of voids. \texttt{VIDE} has been widely used in recent cosmological analysis (e.g. \cite{Sutter2012a,Pisani2014a,Sutter2014c,Hamaus2014b,Hamaus2016,Hamaus2017,Pollina2017}) and embeds the \texttt{ZOBOV} code \citep{Neyrinck2008}.

With \texttt{VIDE} we define the void radius as:
\begin{equation}
R_{\rm{V}}\equiv\bigg( \frac{3}{4\pi}V\bigg)^{1/3}
\end{equation}
where the volume $V$ is the total volume of all the Voronoi cells composing the void (following \texttt{VIDE}'s convention). It is important to notice that \texttt{VIDE} is able to find voids regardless of the shape, so it is particularly adapted to correctly capture the non-spherical feature of voids.

\section{Robustness to Volume and Resolution Effects}
\label{sec:vol_res}

\begin{figure}
\begin{center}
\includegraphics[width=0.5\textwidth]{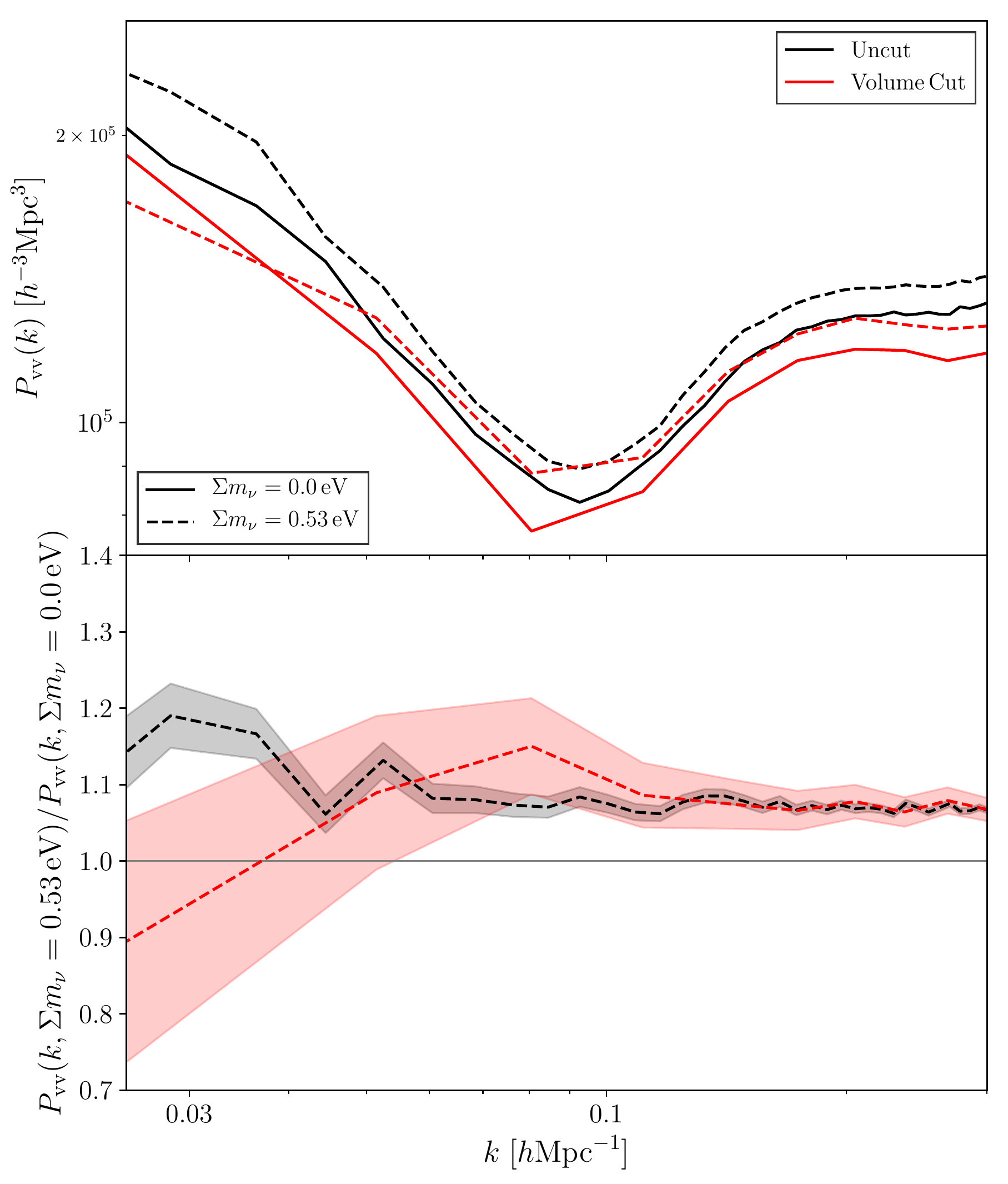}
\caption{Void-void power spectra for voids derived from the {`low-res'} halo distribution. Black spectra correspond to the original uncut {`low-res'} simulation, while red spectra correspond to the volume cut {`low-res'} simulation. Solid lines correspond to $\sum m_\nu =0.0\,\mathrm{eV}$ while dashed lines correspond to $\sum m_\nu =0.53\,\mathrm{eV}$. The bottom panel shows the ratio, with respect to the massless neutrino case, for the uncut and volume cut simulations. The volume cut and uncut simulations are equivalent within uncertainties, so the volume differences between {`low-res'} and {`high-res'} do not induce the inversion.}
\label{fig:vcut}
\end{center}
\end{figure}

To further investigate the inversion described in the main text, we compare results we find with the \texttt{DEMNUni} {(`low-res')} simulations to the smaller but highly resolved \texttt{MassiveNuS} {(`high-res')} simulations described in \S\ref{sec:sim+nbody}.

The main differences between the two simulations are their volume and resolution. Thus, comparing the void behavior in these simulations allows us to check if the inversion in the void abundance and power spectra is a volume and/or resolution artifact or physical in nature.

\subsection{Testing the effect of volume}

Simulation volume can affect the number and size of voids: a simulation with an insufficiently large volume could miss large voids, and if the tracer density is held constant, reducing the simulation volume will decrease the number of voids found, eventually increasing the uncertainties so much that trends become indiscernible. It is therefore important to probe if the volumes of the simulations we use have an effect on our results.

In \autoref{fig:vcut} we plot the {`low-res'} void-void power spectra after cutting the volume of the simulation to match that of the {`high-res'} simulation. We included voids with $x,\,y,$ and $z$ positions $0-512\,h^{-1}\mathrm{Mpc}$ of the origin and removed all others to produce the volume cut catalog.

Cutting the simulation volume maintains the overall shape of the void auto-correlation power spectrum. The elbow near $k\approx 10^{-1}\,h\mathrm{Mpc}^{-1}$ is still present, as is the rise to the left of the elbow. The scales probed by the volume cut simulation are smaller, so the power spectrum spans from only $k=10^{-2}\,h\mathrm{Mpc}^{-1}$ to higher $k$ for which the DEMNUni mass resolution becomes less reliable. For this reason, bins and uncertainties are larger for $k\lesssim 10^{-1}\,h\mathrm{Mpc}^{-1}$ in the volume cut simulation than in the original version.

Since increasing $\sum m_\nu$ still boosts the overall power in the volume cut {`low-res'} simulation, we conclude that the size of the {`low-res'} and {`high-res'} simulations does not influence the inversion behavior we observe.

\subsection{Testing the effect of halo density}
\label{sec:density}

\begin{figure}
\begin{center}
\includegraphics[width=0.5\textwidth]{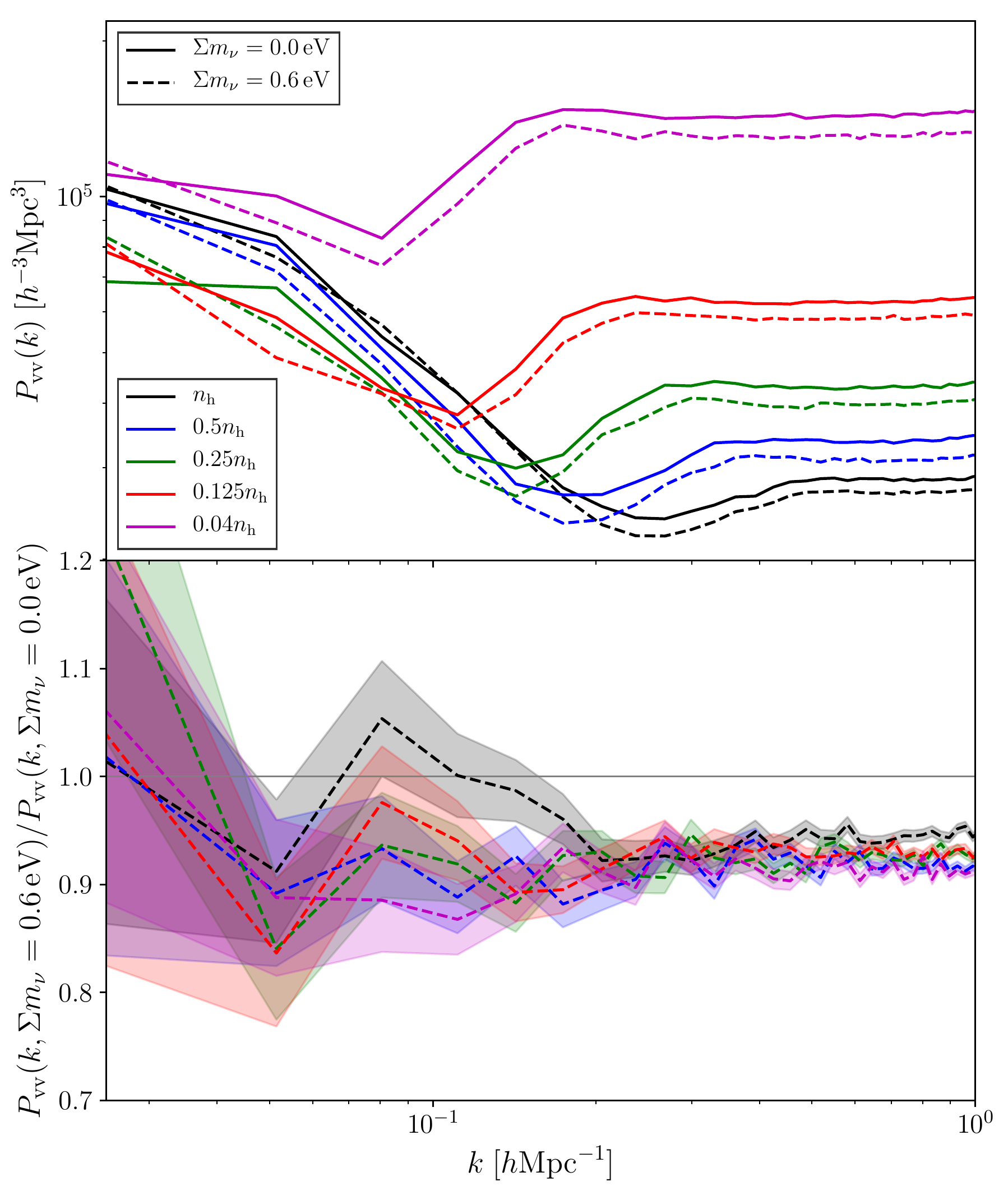}
\caption{Void-void power spectra for the {`high-res'} simulations with different density cuts for voids traced by the halo distribution. Colors denote the tracer density cut of the simulation, where $n_\mathrm{h}$ is the original halo density. The tracer density $0.04n_\mathrm{h}$ corresponds to the halo density for the $M \geq 5 \times 10^{12}\,h^{-1}M_\odot$ mass threshold for the massless neutrino case. Dashed and solid lines denote the values of $\sum m_\nu$ as described in the legend. The bottom panel shows the power spectra ratio with respect to the massless neutrino case, for each density cut simulation. The tracer density does not cause the inversion.}
\label{fig:PvvJia_cuts}
\end{center}
\end{figure}

\begin{figure}
\begin{center}
\includegraphics[width=0.5\textwidth]{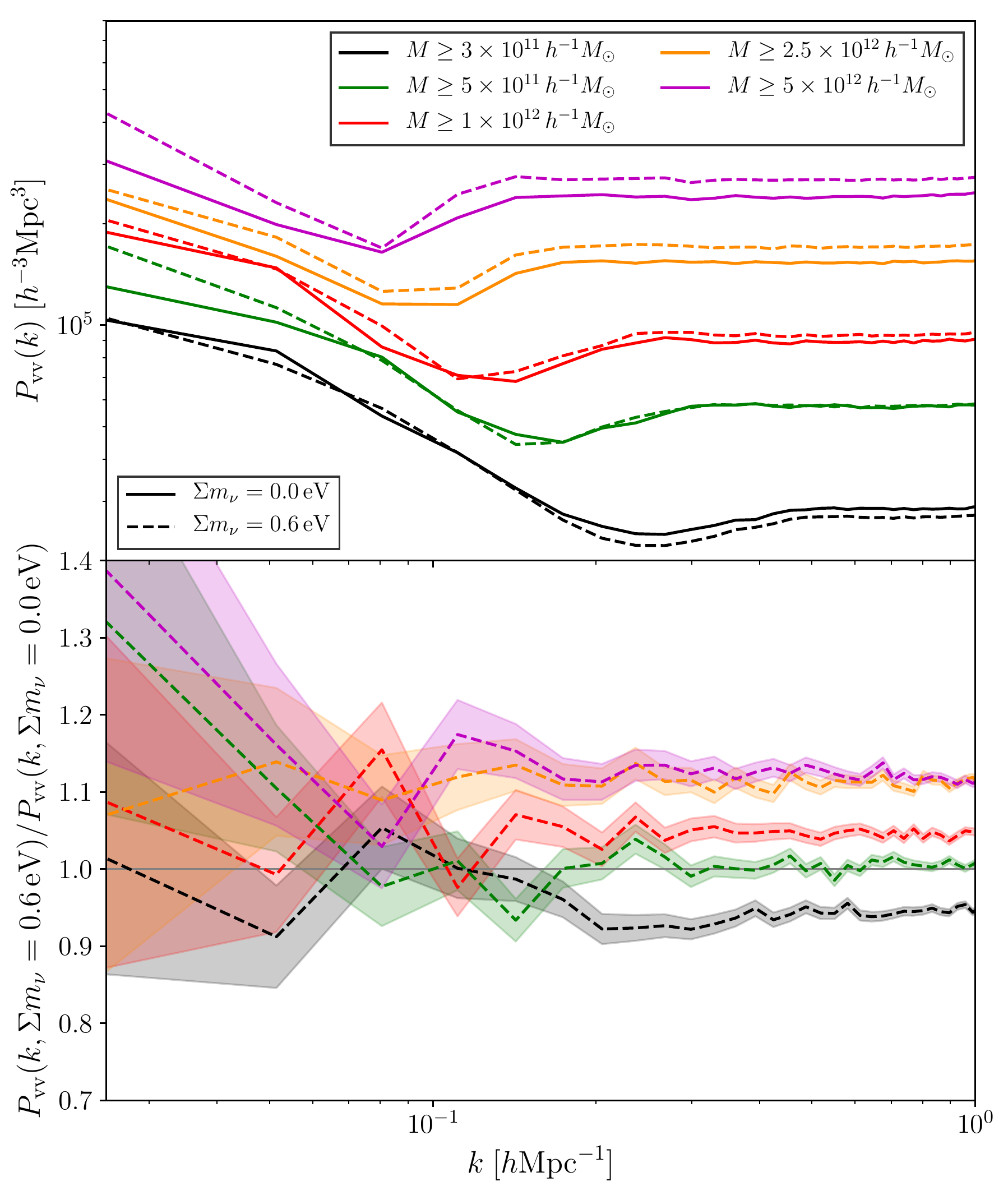}
\caption{Void-void power spectra for the {`high-res'} simulations with different halo mass thresholds to illustrate the effects of the effective halo bias. Colors denote the mass threshold of the simulation, where black is the original mass resolution. Dashed and solid lines denote the sum of neutrino masses used. The bottom panel shows the power spectra ratio between different $\sum m_\nu$ for each halo mass threshold. As the mass threshold increases there is an inversion effect due to a larger effective halo bias and a smaller total number of voids.}
\label{fig:PvvJia_biascuts}
\end{center}
\end{figure}

To probe how halo density affects the inversion, we randomly subsample the {`high-res'} simulation. We plot the void-void power spectra for different halo densities in \autoref{fig:PvvJia_cuts}. Decreasing the halo density shifts the elbow towards large scales because the average void radius increases, and so the exclusion scale increases. Small scales increase in power due to the dependence of shot noise on tracer density \citep{Hamaus2014a}.

While decreasing the tracer density in the {`high-res'} simulation boosts the power, especially at small scales, it does not induce the $\sum m_\nu$ inversion effect. This is in stark contrast to changing the minimum halo mass (see \autoref{fig:PvvJia_biascuts}), which induces an inversion effect as the threshold halo mass increases, increasing the effective halo bias, decreasing the total number of voids, and increasing the average void radius. This suggests physical characteristics of halos induce the inversion effect, justifying the paper's focus on the effective halo bias.

\section{MassiveNuS {(`high-res')} Void Abundance}
\label{sec:massivenus_abundance}

\begin{figure}
\begin{center}
\includegraphics[width=0.5\textwidth]{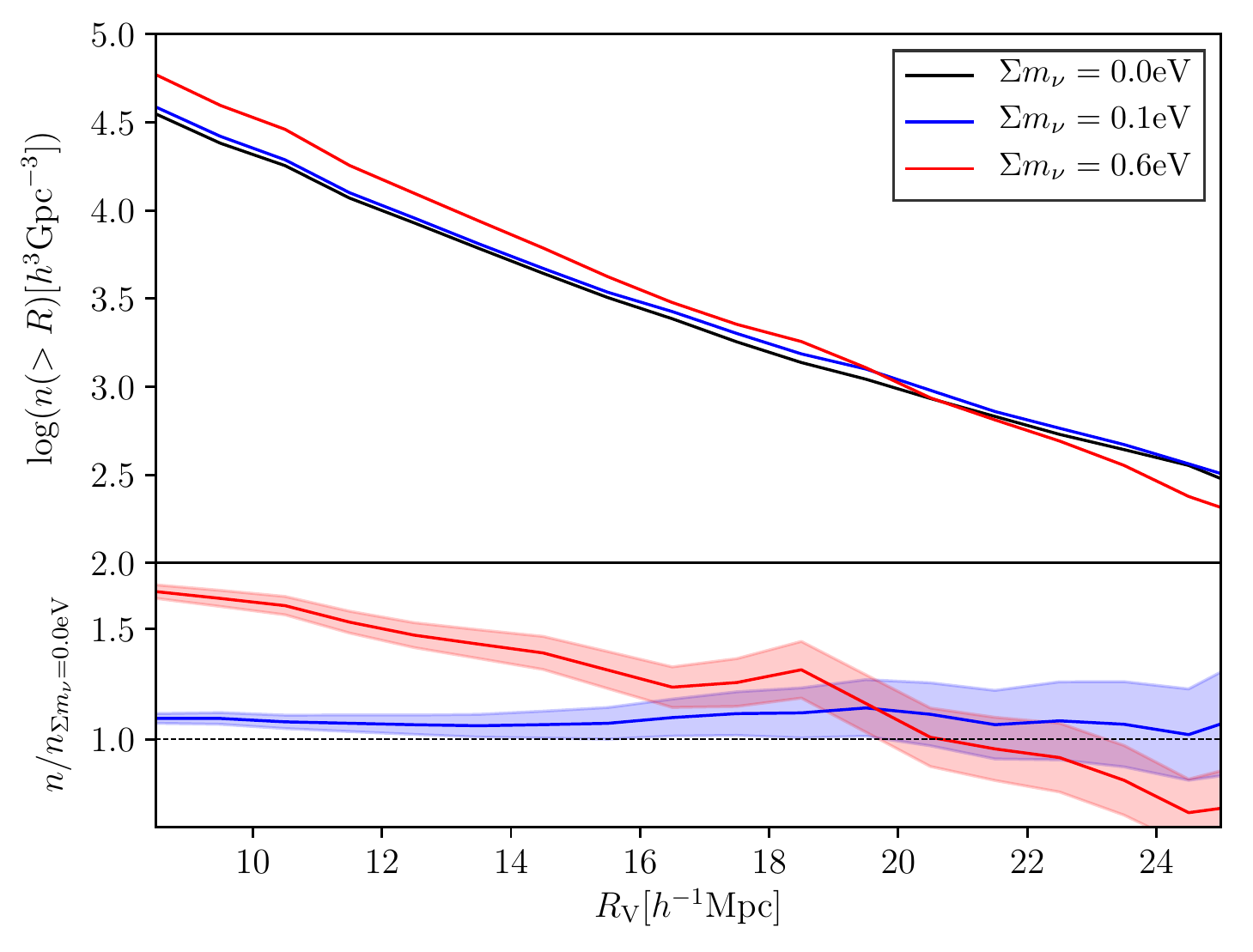}
\caption{Void abundance in the CDM field of the {`high-res'} simulation. Colors denote the sum of neutrino masses used in each simulation. The bottom panel shows the ratio between void number density with uncertainties for the different $\sum m_\nu$ values and the number density in the massless neutrino case. Increasing $\sum m_\nu$ increases the number of small voids and decreases the number of large voids}.
\label{fig:CDM_MassiveNuS_abundance}
\end{center}
\end{figure}

\begin{figure}
\begin{center}
\includegraphics[width=0.5\textwidth]{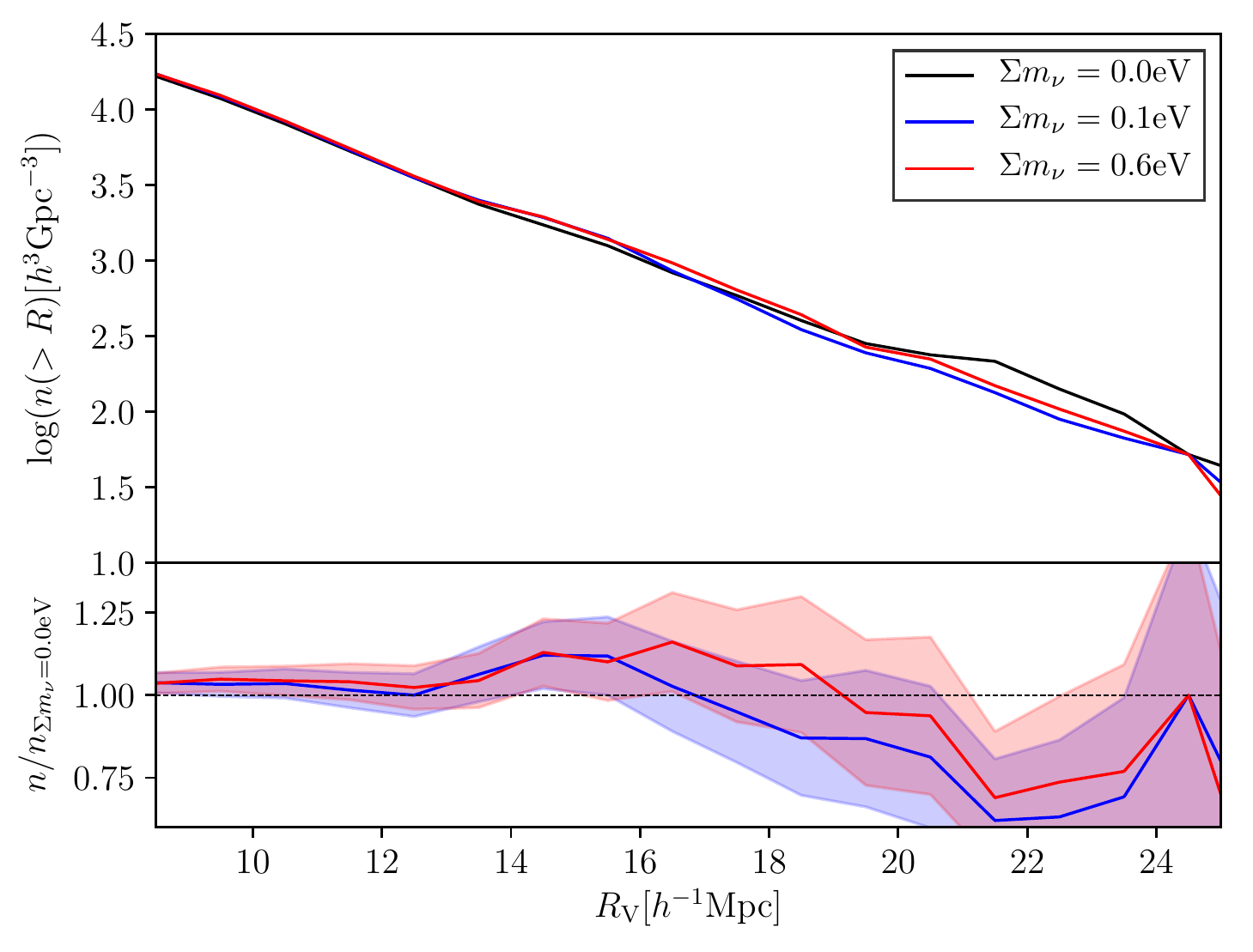}
\caption{Void abundance in the halo field of the {`high-res'} simulation. Colors denote the sum of neutrino masses used in each simulation. The bottom panel shows the ratio between void number densities (with uncertainties) for different $\sum m_\nu$ values and the number density in the massless neutrino case. Nonzero $\sum m_\nu$ appears to increase the number of small voids and decrease the number of large voids relative to the massless case, in contrast to the {`low-res'} abundance for voids traced by halos.}
\label{fig:halo_MassiveNuS_abundance}
\end{center}
\end{figure}

In \autoref{fig:CDM_MassiveNuS_abundance} and \autoref{fig:halo_MassiveNuS_abundance} we show the {`high-res'} abundances for the voids seen in the CDM field and the voids seen in the halo distribution, respectively. Uncertainties are large in \autoref{fig:halo_MassiveNuS_abundance} due to the number of voids, making it difficult to definitively see clear trends for the different $\sum m_\nu$. However, for all $\sum m_\nu$, there are more small voids and less large voids relative to the massless case for voids seen in the halo field. Thus, it appears that abundances for voids seen in both the CDM field and the halo field are consistent with an increased number of small voids and decreased number of large voids as $\sum m_\nu$ increases. This is in contrast to the {`low-res'} abundance plots, which show clear opposite trends for the 2 tracer fields.




\bibliographystyle{mnras}
\bibliography{Voids} 




\bsp	
\label{lastpage}
\end{document}